\documentclass[aps,pra,amsmath,amssymb,amsfonts,superscriptaddress,reprint,twocolumn]{revtex4-2}
\usepackage[colorlinks=true, citecolor=blue, urlcolor=blue ]{hyperref}
\usepackage{epsfig,newlfont,amssymb,amsfonts,amsmath,bm,subfigure,palatino,mathtools,amsthm,braket,soul,enumitem,color,graphics,graphicx,times,physics,makecell,multirow,float}
\usepackage[normalem]{ulem}
\usepackage[ruled,vlined]{algorithm2e}
\usepackage{mathrsfs}

\usepackage{multirow}
\usepackage{braket}

\newcolumntype{P}[1]{>{\centering\arraybackslash}p{#1}}
\graphicspath{{Figures/}{Dynamics/}}
\usepackage{xcolor}

\begin{document}
\title{
Efficient Parameter Estimation through Quantum Extreme Learning
}
\author{Dingwei Zhao}

\affiliation{Institute of Fundamental and Frontier Sciences,
University of Electronic Science and Technology of China, Chengdu 611731, China}

\author{Srijon Ghosh}
\affiliation{Institute of Fundamental and Frontier Sciences,
University of Electronic Science and Technology of China, Chengdu 611731, China}
\affiliation{Key Laboratory of Quantum Physics and Photonic Quantum Information, Ministry of Education,
University of Electronic Science and Technology of China, Chengdu 611731, China}

\author{Victor Montenegro}
\affiliation{College of Computing and Mathematical Sciences, Department of Applied Mathematics and Sciences, Khalifa University of Science and Technology, 127788 Abu Dhabi, United Arab Emirates}

\affiliation{Institute of Fundamental and Frontier Sciences, University of Electronic Science and Technology of China, Chengdu 611731, China}

\affiliation{Key Laboratory of Quantum Physics and Photonic Quantum Information, Ministry of Education, University of Electronic Science and Technology of China, Chengdu 611731, China}

\author{Matteo G. A. Paris}
\affiliation{Dipartimento di Fisica {\it "Aldo Pontremoli"}, Universit\`a degli Studi di Milano, I-20133 Milano, Italia}

\author{Marco G. Genoni}
\affiliation{Dipartimento di Fisica {\it "Aldo Pontremoli"}, Universit\`a degli Studi di Milano, I-20133 Milano, Italia}

\author{Abolfazl Bayat}
\email{abolfazl.bayat@uestc.edu.cn}
\affiliation{Institute of Fundamental and Frontier Sciences,
University of Electronic Science and Technology of China, Chengdu 611731, China}
\affiliation{Key Laboratory of Quantum Physics and Photonic Quantum Information, Ministry of Education,
University of Electronic Science and Technology of China, Chengdu 611731, China}
\affiliation{Shimmer Center, Tianfu Jiangxi Laboratory, Chengdu 641419, China}

\begin{abstract}
Quantum sensing protocols rely on the efficient processing of measurement outcomes to infer unknown physical parameters. Conventional optimal estimators, such as Bayesian inference and maximum likelihood estimation, require to construct a likelihood function whose efficient classical construction requires full knowledge of the dynamics, which may practically be infeasible.  Model-free experimental construction of the likelihood is also very resource intensive. Here, we propose a quantum extreme learning-based framework for quantum parameter estimation which avoids the explicit construction of the likelihood function. We demonstrate the effectiveness of this approach using the transverse-field Ising model for estimating the external transverse field. Our results show that high estimation accuracy can be achieved using measurements at only a small number of training points. We further develop an interrogation-time optimization strategy that 
minimizes the estimation inaccuracy while keeping sampling fluctuations below a prescribed threshold. We further extend the framework to a two-time measurement protocol, in which measurement outcomes obtained at two distinct times are combined to enrich the feature space, thereby significantly enhancing the estimation accuracy.  Our analysis shows that the estimation precision improves monotonically with the number of training points, while increasing the system size generally enhances the sensing performance, albeit non-monotonically. Finally, the protocol is shown to be robust against dephasing, making it suitable for near-term quantum devices. 
\end{abstract}

\maketitle

\section{Introduction}

Quantum sensing has emerged as one of the most promising applications of quantum technologies~\cite{degen2017quantum,pezze2018quantum,braun2018quantum,ye2024essay, ghosh2026journey}, offering unprecedented precision in estimation of physical parameters through exploiting quantum features such as entanglement~\cite{giovannetti2004quantum,giovannetti2011advances,giovannetti2006quantum}, squeezing~\cite{pezze2008mach} and criticality~\cite{montenegro2025quantum,mihailescu2026critical}. A typical quantum sensing protocol involves four essential steps: (i) preparation of an appropriate quantum probe; (ii)  encoding of the unknown parameter through the dynamics of the probe; (iii)  measurement of a suitable observable on the evolved probe; and (iv) feeding the measurement outcomes into an estimator to infer the unknown parameter. Ideally, for a given probe, one has to optimize both the measurement and the estimator to achieve the best possible performance. However, in many practical scenarios, measurement operations are constrained by experimental limitations and are often restricted to specific local measurement bases. In contrast, estimators offer greater flexibility, as they are part of the classical post-processing of the measurement data.

Bayesian inference and its closely related maximum likelihood estimation are proven to be the optimal estimators in the asymptotic limit of large measurement samples~\cite{hradil1996quantum,pezze2007phase,olivares2009bayesian, Brivio2010, Genoni2012, Blandino2012, le2012asymptotic,rubio2019quantum, Gianani2020}. These estimators aim to infer the unknown parameter $h$ by constructing a probability distribution, known as the posterior,  conditioned on the observed measurement outcomes. Based on Bayes' theorem, the posterior probability distribution $p(h|\mathrm{data})$ can be written as $p(h|\mathrm{data}) \propto p(\mathrm{data}|h)p(h)$. Here, $p(h)$ is the prior distribution, which quantifies the information available about the unknown parameter before any measurements are performed, and  $p(\mathrm{data}|h)$ is the likelihood function, which describes the probability of observing the measured data given that the unknown parameter takes the value $h$.
Efficient construction of the likelihood function on a classical computer requires having full knowledge about the probe and its evolution.  Moreover, in practice, experimental imperfections often prevent us from having such complete knowledge, as, for instance, Hamiltonian parameters or decoherence effects may not be precisely known. Even if full knowledge of the evolution exists, classical simulation of large quantum systems is in general limited by curse of dimensionality.
A possible way to circumvent the classical construction of the likelihood  is to adopt a 
model-free approach, in which the likelihood function is constructed experimentally without any assumption about the details of the evolution. However, this approach is often highly resource-intensive, as it requires a large number of measurement samples to accurately estimate the probability distributions at different, carefully controlled values of $h$ across the parameter range of interest. This typically involves sampling over a sufficiently fine parameter grid to achieve the desired accuracy (see, e.g., Refs.~\cite{yu2025experimental,li2026non}). 
This naturally raises an important question: can one develop a model-free and resource-efficient estimator that can infer unknown physical parameters without explicitly constructing the likelihood function?

A possible solution to address the above question is to exploit machine-learning techniques, where the relationship between the unknown parameter and the corresponding measurement outcomes can be learned directly from a set of training data \cite{nolan2021machine,nolan2021frequentist,huang2025quantum,xiao2022parameter,schuff2020improving,costa2021benchmarking}, thereby avoiding the explicit construction of the likelihood function. In this context, quantum reservoir computing~\cite{mujal2021opportunities} and one of its variants, Quantum Extreme Learning (QEL)~\cite{innocenti2023potential}, have recently emerged as promising frameworks that exploit the complex dynamics of quantum systems for solving various tasks, including entanglement detection~\cite{zia2025quantum}, quantum state characterization~\cite{garcia2023scalable}, time-series analysis~\cite{garcia2023scalable,mujal2023time,das2026quantum,li2026quantum,kobayashi2024feedback,Vetrano_2025,liu2026time,sannia2026temporalinformationprocessing4500qubit,xia2022reservoir,settino2025memory}, solving nonlinear difference equations~\cite{henaff2024optical,hou2026high,llodra2025quantum}, chaos characterization~\cite{li2026quantum,sharifian2026quantum} and image processing~\cite{zhang2026robust}. In these reservoir-based learning methods, the input information is encoded into a quantum system, whose dynamics generate a high-dimensional representation of the input. The resulting information is extracted through measurements of selected observables and processed using a trainable classical readout layer \cite{mujal2021opportunities,innocenti2023potential}. Unlike conventional variational quantum algorithms~\cite{cerezo2021variational}, which require optimizing a large number of parameters governing the quantum dynamics, these reservoir-based computing methods rely on fixed reservoir dynamics, with training restricted to the classical readout. Thus, one may wonder whether QEL can also be used as an efficient estimator that can directly infer the unknown physical parameters through measurement data.

In this work, we propose a QEL-based framework for estimating an unknown parameter encoded in the reservoir Hamiltonian. 
Unlike conventional QEL approaches, in which the quantum system serves as a processor for externally supplied information, our protocol keeps the input state fixed and instead exploits the parameter-dependent system dynamics to infer the unknown parameter. We demonstrate the effectiveness of our approach using the transverse-field Ising model, where the external transverse field is the parameter to be estimated. Our results show that it is possible to achieve high sensing precision using measurements at only a small number of controlled values of the parameter as the training points. Furthermore, we develop an interrogation-time optimization procedure that minimizes the estimation error while ensuring that sample fluctuations remain below a prescribed threshold. This optimization reveals that the protocol efficiently reaches high accuracy at a fairly fast evolution time. We further extend the QEL framework to a two-time measurement protocol, in which measurement outcomes at two different times are combined to enlarge the feature space, thereby significantly enhancing the estimation precision. Moreover, our results show that the estimation precision improves monotonically with the number of training points. Increasing the system size also generally enhances the estimation precision, although the improvement is not strictly monotonic. Finally, the protocol is shown to be robust against dephasing and thus can be used in existing quantum devices.  \\

The paper is organized as follows: In Sec. \ref{method}, we begin by introducing the protocol for estimating the parameter using QEL. Then, in Sec. \ref{model}, we consider the transverse-field Ising spin chain to demonstrate the efficiency of the proposed scheme. We discuss the numerical findings in Sec. \ref{result}. The robustness of the method in the presence of the environment is addressed in Sec. \ref{decoherence}. Finally, we conclude in Sec. \ref{conclusion}.

\section{Quantum Extreme Learning-Based Sensing}
\label{method}

QEL has emerged as a paradigm for quantum machine learning with various applications~\cite{zia2025quantum,garcia2023scalable,mujal2023time,das2026quantum,li2026quantum,kobayashi2024feedback,Vetrano_2025,liu2026time,sannia2026temporalinformationprocessing4500qubit,xia2022reservoir,settino2025memory,henaff2024optical,hou2026high,llodra2025quantum,li2026quantum,sharifian2026quantum,zhang2026robust}. Conventionally, information is encoded in the initial state of the system which is subsequently evolved by a fixed quantum reservoir~\footnote{Note that, in QEL, quantum reservoir refers to the quantum system used to encode and process the input data, rather than to an environmental bath.}. After the evolution, some local operators are measured and their averages are used for estimating the parameter of interest~\cite{mujal2021opportunities,innocenti2023potential}. In contrast, our proposed approach considers a quantum reservoir whose Hamiltonian itself depends on an unknown parameter that we aim to estimate. Consequently, the parameter is not encoded in the initial state but in the reservoir dynamics. By measuring suitable observables of the evolved state, we extract information about the unknown parameter and use a trained classical readout to estimate its value. 

Consider a reservoir whose Hamiltonian $H_{h}$  depends on an unknown parameter $h$. Our objective is to estimate this parameter using QEL. The proposed protocol consists of four steps. (i) \textit{Preparation of initial state: } To encode the parameter, we consider an initial state $|\psi(0)\rangle$ that  should not be an eigenstate of $H_{h}$. (ii) \textit{Encoding via evolution: } At first, let us assume that the system is a closed system and the parameter is encoded in the initial state via unitary evolution governed by the unitary $U_{h} = \exp(-i H_{h} t)$. The evolved state reads as $|\psi_{h}(t)\rangle = U_{h} |\psi(0)\rangle$, where $t$ is the total time of evolution. 
(iii) \textit{Readout:} After the system evolves for a time \(t\), the expectation values of a set of $L$ local observables \(\{\mathcal{O}_j\}\), $j=1,2,\ldots,L$, are computed. The expectation values are collected into a vector with $L$ elements as
\begin{equation}\label{eq:expectaions_general}
\mathbf{V}(h,t)=\left(\langle \mathcal{O}_1\rangle,\langle \mathcal{O}_2\rangle\ldots,
\langle \mathcal{O}_L\rangle\right)^T
\end{equation}
where $\langle \mathcal{O}_j\rangle=\bra{\psi_h(t)}\mathcal{O}_j\ket{\psi_h(t)}$ represents the average of the operator $\mathcal{O}_j$ with respect to $\ket{\psi_h(t)}$. (iv) \textit{Parameter estimation:}
We assume that the parameter $h$ can be estimated from the measurement vector $\mathbf{V}(h,t)$ through linear regression $h_{\mathrm est}{=}\mathbf{W}(t)\mathbf{V}(h,t)$, where $\mathbf{W}(t)$ is a trainable weight matrix of size $1{\times} L$. 

The key point in the above protocol is to find the optimal weight matrix $\mathbf{W}$. 
To train the matrix $\mathbf{W}$,  we choose $N_{\mathrm{tr}}$
samples of the parameter $h$ and make a  vector $\boldsymbol{h}_{\mathrm{tr}}{=}(h_1,h_2,\cdots,h_{N_{\mathrm{tr}}})^T$ from a specified
interval $h{\in}[h_{\min},h_{\max}]$.
For each of these training points, the corresponding feature vector
$\mathbf{V}(h,t)$ is constructed from the expectation
values of the measured observables. These feature vectors
are then arranged column-wise to form the training matrix
\begin{equation}
\mathbf{M}_{\mathrm{tr}} =
\left[
\mathbf{V}(h_1,t),
\mathbf{V}(h_2,t),
\ldots,
\mathbf{V}(h_{N_{\mathrm{tr}}},t)
\right].
\end{equation}
The task of parameter estimation is then reduced to
determining an optimal set of classical readout weights
$\mathbf{W}$ that maps the measured features to the
corresponding parameter values. To this end, we employ
ridge regression and minimize the loss
function, defined via the Frobenius norm
\begin{equation}
\mathcal{L}(\mathbf{W}) =
\left\|
\mathbf{W}\mathbf{M}_{\mathrm{tr}}
-\boldsymbol{h}_{\mathrm{tr}}
\right\|_F^2,
\end{equation}
Minimizing this loss function with respect to matrix $\mathbf{W}$ yields the optimal
readout weights \cite{fujii2017, ghosh2021, kaito2024}
\begin{equation}
\mathbf{W}^{\mathrm{opt}}(t) =
\boldsymbol{h}_{\mathrm{tr}}
\mathbf{M}_{\mathrm{tr}}^{T}
\left(
\mathbf{M}_{\mathrm{tr}}
\mathbf{M}_{\mathrm{tr}}^{T}
+\epsilon\mathbf{I}
\right)^{-1}.
\end{equation}
Here, $\epsilon>0$ is a small regularization parameter to ensure the matrix remains non-singular. For all numerical calculations, the regularization parameter is fixed at $\epsilon {=} 10^{-8}$.

Once the training is completed, the optimal weights are kept fixed. To see the effectiveness of the protocol, we should test its performance on unseen values of $h$ which are not included in the training set $\boldsymbol{h}_{\mathrm{tr}}$. Thus, we consider a test set $\boldsymbol{h} {=} (h_{1}, h_{2}, \ldots, h_{N_{\mathrm{test}}})^T$ and construct its corresponding matrix from the measurement outcomes $\mathbf{M}_{\mathrm{test}}(\boldsymbol{h},t)$. The estimated value is obtained as
\begin{equation}
\boldsymbol{h}_{\mathrm{est}} (t) =
\mathbf{W}^{\mathrm{opt}}(t)
\mathbf{M}_{\mathrm{test}}(\boldsymbol{h},t).
\end{equation}
where, $\boldsymbol{h}_{est} (t) = (\widehat h_{1} (t), \widehat h_{2} (t), \ldots, \widehat h_{N_{\mathrm{test}}}(t))^T$. To quantify the accuracy of the estimation protocol, we calculate the root-mean-square error (RMSE), defined as
\begin{equation}
\mathrm{RMSE}(t)
=
\sqrt{
\frac{1}{N_{\mathrm{test}}}
\sum_{k=1}^{N_{\mathrm{test}}}
\left[
\widehat{h}_{k}(t)-h_{k}
\right]^2
}.
\label{eq:rmse}
\end{equation}
Note that RMSE(t) is also a function of the total number of qubits $N$, and the number of training points $N_{\mathrm{tr}}$.

It is worth emphasizing that while RMSE provides a good figure of merit for quantifying the accuracy of our sensing protocol,  it requires to measure expectation values of several local operators, see Eq.~(\ref{eq:expectaions_general}). Those expectation values, however, must be interpreted in the limit of infinite samplings. For any finite number of samples per qubit $M$, the expectation value of the local operator $\mathcal{O}_j$  can only be estimated within a precision range determined by statistical fluctuations given by $\sqrt{[\langle\mathcal{O}_j^2\rangle-\langle\mathcal{O}_j\rangle^2]/M}$. Consequently, a protocol exhibiting a small RMSE may still require a large number of measurements to achieve the desired estimation precision. Therefore, to provide a more comprehensive assessment of the proposed scheme, it is important to investigate how these measurement fluctuations propagate into the estimated parameter. To quantify this effect, we introduce the standard deviation \(\sigma\) of the estimated parameter for a single-shot measurement, i.e. $M{=}1$, defined as
\begin{equation}
    \sigma(h,t)=\sqrt{
\sum_{j=1}^{2N}
\mathbf{W}_j^2(t)
\left[
\langle\mathcal{O}_j^2\rangle-\langle\mathcal{O}_j\rangle^2
\right]},
\label{eq:sigma_general}
\end{equation}
which characterizes its statistical uncertainty arising from quantum measurement fluctuations. For finite $M$, the fluctuations can always be estimated from this single-shot quantity as $\sigma(h,t)/\sqrt{M}$. 
Note that in the above equation, we assume that every local operator $\mathcal{O}_j$ is measured independently. If one measures those operators simultaneously, then  covariances of multi-body operators have to be included in the computation of the above standard deviation. For simplicity and without loss of generality, we restrict ourselves to Eq.~(\ref{eq:sigma_general}). Thus, to make sure that the statistical  error does not become larger than a specified threshold, one has to choose the  samples size per qubit $M$ accordingly. The total measurement cost of the protocol then would be $ML$.



In order to get a quantity independent of the unknown parameter $h$ one can compute the average 
of $\sigma$ with respect to $h$ as
\begin{equation}
\overline{\sigma}(t)=
\frac{1}{N_{\mathrm{test}}}
\sum_{k=1}^{N_{\mathrm{test}}}
\sigma(h_k,t).
\label{average_sigma}
\end{equation}
It is worth emphasizing that, in the quantum sensing community, sensing ``accuracy" refers to the deviation from the true parameter value
and is thus quantified by RMSE. On the other hand, sensing ``precision" refers to the statistical uncertainty for a finite number of samples and is quantified by $\sigma(h,t)$. If this quantity remains almost independent of $h$, as we will see is the case in our model, then one can use $\overline{\sigma}(t)$ as a measure of precision. Throughout this paper we use the same terminology and refer to RMSE as the accuracy and $\overline{\sigma}(t)$ as precision.

\begin{figure*}[t]
    \includegraphics[width=0.9\textwidth]{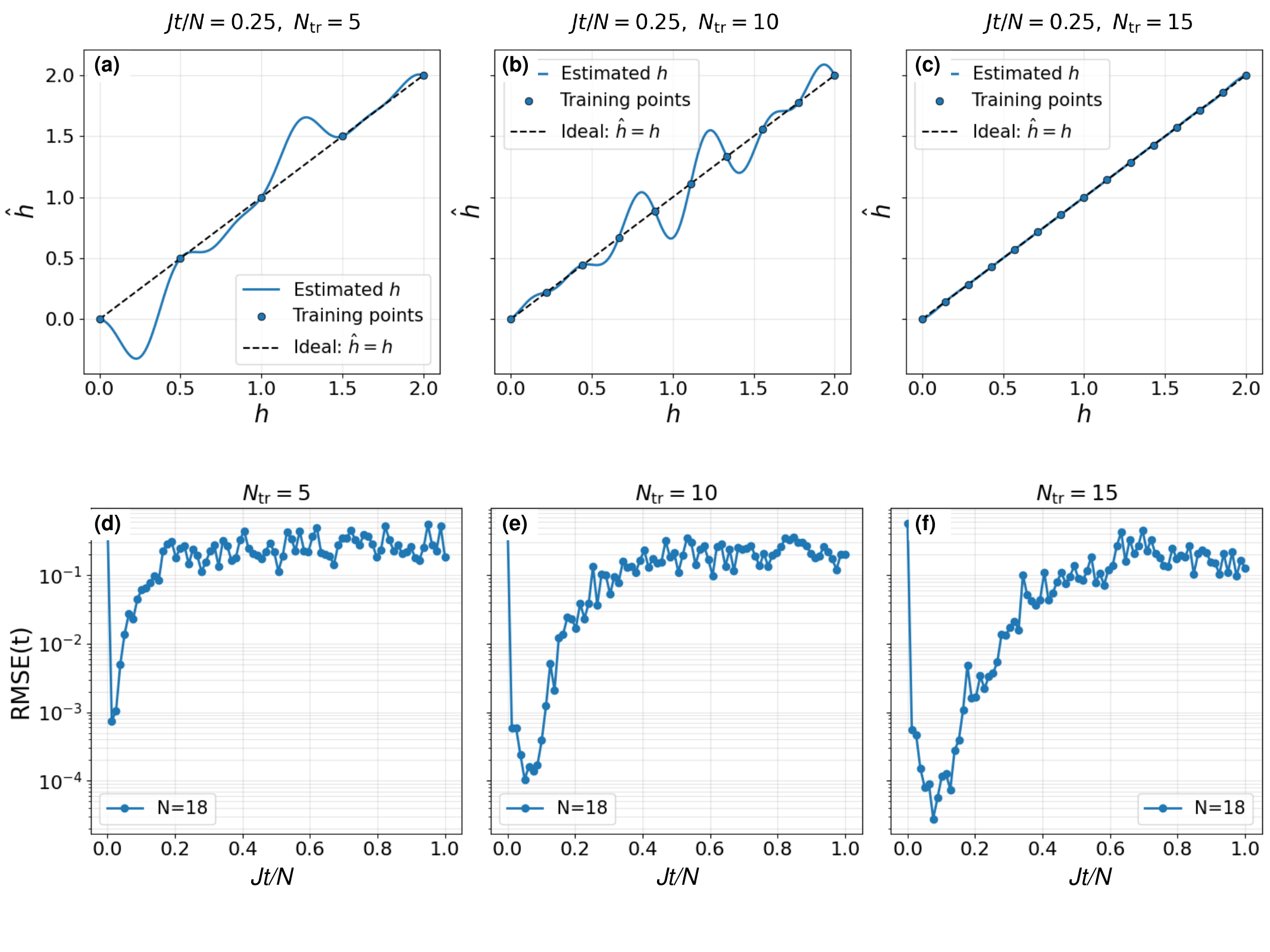}
    \caption{Comparison of parameter estimation accuracy for different numbers of training points $N_{\mathrm{tr}}$. Upper panel: Estimated parameter $(\widehat h)$ as a function of its true value $(h)$ for different training-set sizes $N_{\mathrm{tr}}$, at a fixed evolution time $t=0.25N/J$. The black dashed line represents perfect estimation $\widehat h=h$, while the blue circles indicate the training points. Lower panel: Root-mean-square error (RMSE) as a function of the normalized evolution time $(J t/N)$ for the corresponding training-set sizes. Here $N = 18$ and $h \in [0,2]$. }
    \label{h_real_vs_h_estimated}
\end{figure*}

\section{the Model}
\label{model}

To demonstrate the effectiveness of the proposed scheme, we consider a transverse field Ising chain of size $N$ with open boundary conditions. The Hamiltonian reads as
\begin{equation}
    H =  - J \left[ \sum_{i = 1}^{N-1} {X}_{i} X_{i+1} + h \sum_{i = 1}^{N} {Z}_{i}\right],
    \label{TFIM}
\end{equation}
where $X$ and $Z$ are the Pauli matrices. $J$ is the interaction strength between the nearest-neighbor qubits and $h$ is a dimensionless quantity denoting the strength of the transverse field with respect to the exchange coupling. For all the numerical simulations, we restrict the value of $h \in [0,2]$. We consider an initial state as 
\begin{equation}
|\psi(0)\rangle
=\bigotimes_{j=1}^{N/2}
\left(|0\rangle_{2j-1}\otimes|+\rangle_{2j}\right),
\end{equation}
where $|0\rangle$ and $|+\rangle$ are the eigenbasis of $Z$ and $X$, respectively. The motivation for choosing this initial state is that it is not an eigenstate of the Hamiltonian throughout the entire range of the parameter $h$. Consequently, the state exhibits nontrivial dynamics in both limiting cases, allowing information about the Hamiltonian parameters to be encoded in its time evolution. The transverse-field Ising model is exactly solvable through the Jordan–Wigner transformation \cite{barouch1970}, which maps the spin operators onto fermionic operators. By introducing Majorana operators, the Hamiltonian in Eq.~(\ref{TFIM}) can be expressed in the quadratic form \(H=(i/4)\boldsymbol{\gamma}^{T}A\boldsymbol{\gamma}\), where \(A\) is a real antisymmetric matrix of dimension \(2N\times 2N\). This representation avoids the exponential growth of the Hilbert-space dimension encountered in exact diagonalization and enables efficient numerical simulations for substantially larger system sizes \cite{santoro2024}. The details of the Hamiltonian diagonalization is given in Appendix \ref{diagonalization}.

Let us consider a set of \(N_{\mathrm{tr}}\) training values of the parameter \(h\), denoted by \(\mathscr{H}_{\mathrm{tr}}=(h_1^{\mathrm{tr}},\ldots,h_{N_{\mathrm{tr}}}^{\mathrm{tr}})\). For each training value \(h_i^{\mathrm{tr}}\), the corresponding parameter-encoded state \(|\psi(h_i^{\mathrm{tr}},t)\rangle\) is obtained by evolving the initial state under the unitary operator \(U(h_i^{\mathrm{tr}},t)=\exp[-iH(h_i^{\mathrm{tr}})t]\). For this model, we calculate the feature vector as
\begin{equation}\label{eq:local_expectations}
    \mathbf{V} (h^{tr}_{i}, t) = \left(\langle X_1\rangle, \langle X_2\rangle,  \ldots, \langle X_N\rangle,  \langle Z_1 \rangle, \ldots, \langle Z_N\rangle \right)^T 
\end{equation}
where, $\langle X_i \rangle {=}
\langle \psi(t) | X_i | \psi(t) \rangle$ and $\langle Z_i \rangle {=}
\langle \psi(t) | Z_i | \psi(t) \rangle$. Therefore, the feature vector
$\mathbf{V}(h_i^{\mathrm{tr}},t)$ has dimensions
$2N {\times }1$, namely $L{=}2N$. For large system sizes $N$, these
time-dependent local spin expectation values, which
serve as reservoir features, can be calculated using
the corresponding Majorana correlation functions, see Appendix~\ref{diagonalization} for more details.

\section{Numerical Results}
\label{result}

Having established the training procedure and the transverse Ising model as our reservoir, we now investigate the performance of the proposed quantum extreme learning-based parameter estimation protocol. In Figs.~\ref{h_real_vs_h_estimated}(a)-(c) we plot the estimated value $\widehat h$ as a function of true value $h$ for a system of size $N{=}18$ and fixing time to $t{=}0.25N/J$ for different number of training points $N_{\mathrm{tr}}$.  Here, we consider 200 test values of $h$ in the interval $h \in [0,2]$. As the figure shows, by increasing the number of training points the estimated values of the transverse magnetic field approach their corresponding true values over the entire interval.  Remarkably, even with $N_{\mathrm{tr}} = 15$ training points, the maximum absolute error $\mathrm{MAE} = \max_{1\leq k\leq N_{\mathrm{test}}}|\widehat h_k(t)-h_k |$ is as low as $0.01342$ demonstrating the effectiveness of the proposed estimation protocol (see Table~\ref{table}). 

\begin{table}[t]
\centering
\begin{ruledtabular}
\begin{tabular}{cccc}
$N_{\mathrm{tr}}$ &
MAE &
$h$ &
$\widehat h$ \\
\hline
5  & 0.562189 & 0.251256 & -0.310933 \\
10 & 0.334947 & 1.005025 &  0.670078 \\
15 & 0.013415 & 1.939698 &  1.953114 \\
\end{tabular}
\end{ruledtabular}
\caption{Maximum absolute error in the estimation of the
transverse magnetic field $h$ for different numbers of
training points $N_{\mathrm{tr}}$. The corresponding
true and estimated values of $h$ at which the maximum
error occurs are also reported.}
\label{table}
\end{table}

To see the effect of time on the performance of the protocol, in  Figs.~\ref{h_real_vs_h_estimated}(d)-(f) we plot  RMSE as a function of time for a reservoir of size $N{=}18$ for different number of training points $N_{\mathrm{tr}}$. As expected, the RMSE is maximum at $t{=}0$, since the initial state does not contain any information about the unknown parameter $h$. Interestingly, as the evolution time increases, the RMSE decreases and reaches a minimum at an optimal time $t{=}t^*$, beyond which it increases again and eventually tends to stabilize. Nonetheless, the minimum of RMSE decreases by increasing the training points and the optimal time $t^*$ changes as $N_{\mathrm{tr}}$ varies.  This behavior highlights the crucial role of the evolution time in determining the performance of the proposed sensing protocol. 

\begin{figure}[t]
    \includegraphics[width=\linewidth]{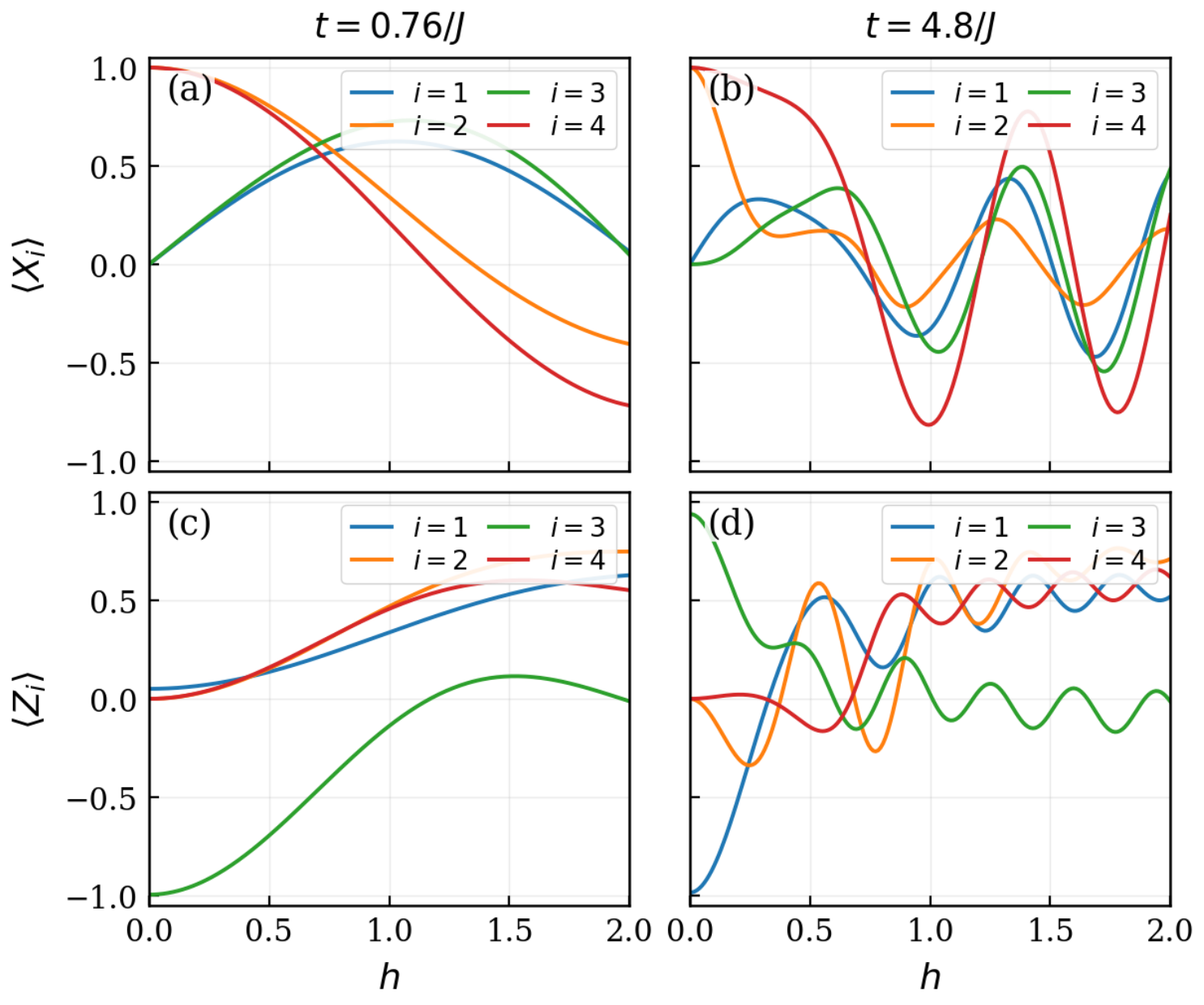}
\caption{Variation of the expectation values of the local observables $X_i$ and $Z_i$ as functions of the transverse field $h$ for a system of $N=4$. The evolution time $t=0.76/J$ corresponds to the minimum root-mean-square error (RMSE), while $t=4.8/J$ represents an intermediate stage of the system's evolution.}
    \label{N4}
\end{figure}

Note that in non-equilibrium quantum sensing protocols in which the probe undergoes a unitary evolution, typically the ultimate precision limit (which assumes optimal measurements), quantified by quantum Fisher information~\cite{paris2009quantum,liu2020quantum,meyer2021fisher}, scales quadratically in time (i.e. $\sim t^2$)~\cite{bixo2007generalized,o2026geometric,zheng2026stabilizing}, whereas our restricted local-Pauli strategy does not.
The suboptimal performance can be attributed to two main factors: (i) our measurements are restricted to local Pauli operators, which generally differ from the optimal measurement basis required to saturate the quantum Fisher information bound; and (ii) we only rely on the expectation values of measurement outcomes, which, although experimentally convenient, capture less information than the full probability distributions of all possible measurement outcomes. Consequently, only a restricted portion of the parameter-dependent information encoded in the evolved state is accessible through the expectation values of local measurements. 
As the system evolves, the reduced density matrices of individual qubits initially accumulate information about the unknown field, as reflected in the decreasing RMSE. Eventually, they equilibrate and their information content saturates, as reflected in the long-time saturation of the RMSE in Figs.~\ref{h_real_vs_h_estimated}(d)-(f). To understand this, in Fig.~\ref{N4}, we plot the expectation values $\langle X_j\rangle$ and $\langle Z_j\rangle$ as functions of $h$ at two different evolution times in a reservoir of size $N{=}4$. At the shorter time, $t=0.76/J$ (see Fig.~\ref{N4} (a) and (c)), corresponding to the minimum RMSE, the expectation values exhibit relatively smooth variations with $h$, providing a more distinguishable relationship between the measurement outcomes and the unknown parameter. In contrast, at the longer time, $t=4.8/J$ (see Fig.~\ref{N4} (b) and (d)), the observables develop pronounced oscillations and non-monotonic behavior, such that different values of $h$ can produce similar local expectation values. This makes the inverse mapping from the measured observables to the transverse field increasingly ambiguous and more difficult to learn from a finite training set. This behavior is directly related to exponential concentration issue which is known to be limiting factor to all quantum reservoir based computing schemes such as QEL~\cite{thanasilp2024exponential, sannia2025exponential,xiong2025fundamental}. Therefore, although longer evolution can enhance the parameter sensitivity of the quantum state, this enhancement does not necessarily translate into improved estimation accuracy when only a limited set of local observables is employed.


\begin{figure*}[t]
    \centering
    \includegraphics[width=0.9\textwidth]{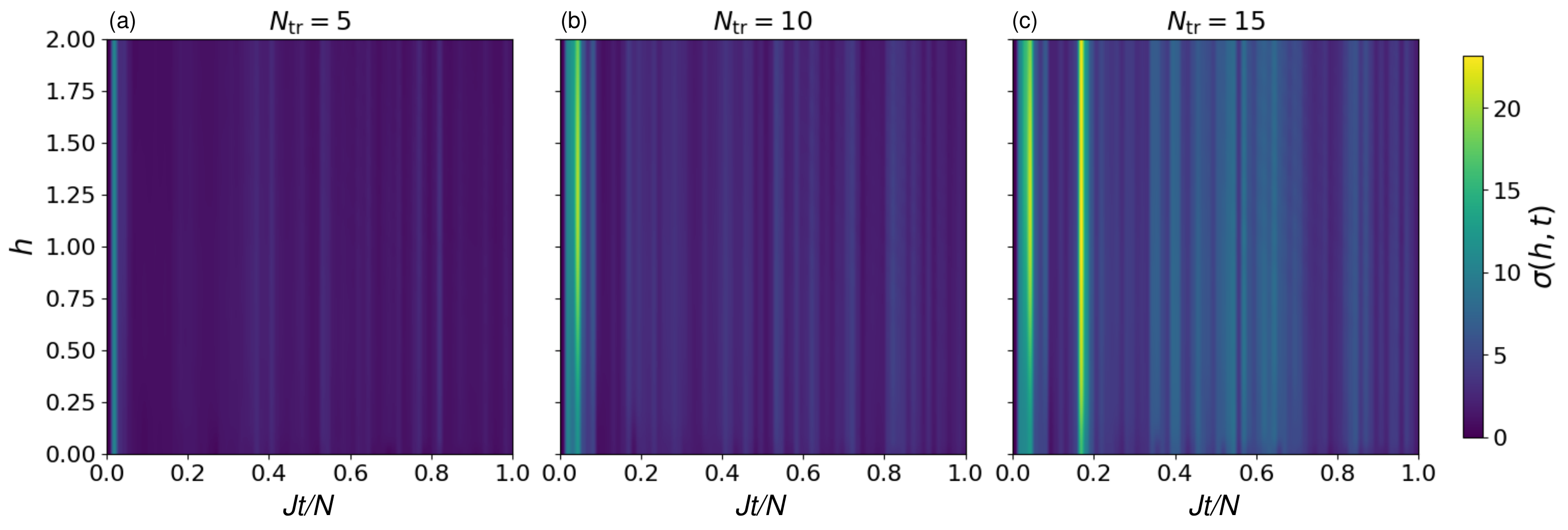}
    \caption {Single-shot measurement uncertainty $\sigma$ as a function of normalized time and real values of $h$. Here $N = 18$. }
    \label{sigma_vs_t}
\end{figure*}

Since, in this protocol, we are mainly concerned about the expectation values of the Pauli operators, i.e. $\mathcal{O}_j{=}X_j$, $ Z_j$,  Eq.~(\ref{eq:sigma_general}) reduces to 
\begin{equation}
    \sigma(h,t)=\sqrt{
\sum_{j=1}^{2N}
\mathbf{W}_j^2(t)
\left[
1-\langle\mathcal{O}_j\rangle^2
\right]},
\end{equation}
where, 
\begin{equation}
\mathcal{O}_j=
\begin{cases}
X_{j},
& 1\leq j\leq N,\\[4pt]
Z_{j-N},
& N+1\leq j\leq 2N.
\end{cases}
\label{reservoir_features}
\end{equation}

In Figs.~\ref{sigma_vs_t}(a)-(c), we plot $\sigma(h,t)$ as a function of parameter $h$ and time $t$ for a reservoir of size $N{=}18$ and different number of training points $N_{\mathrm{tr}}$, respectively. Interestingly, regardless of the number of training points, measurement uncertainty is nearly uniform with respect to $h$ but varies in time. This motivates us to define the average single-shot measurement uncertainty $\overline \sigma(t)$, see Eq.~(\ref{average_sigma}), as the  precision our protocol. 

\begin{figure*}[t]
    \includegraphics[width=0.9\textwidth]{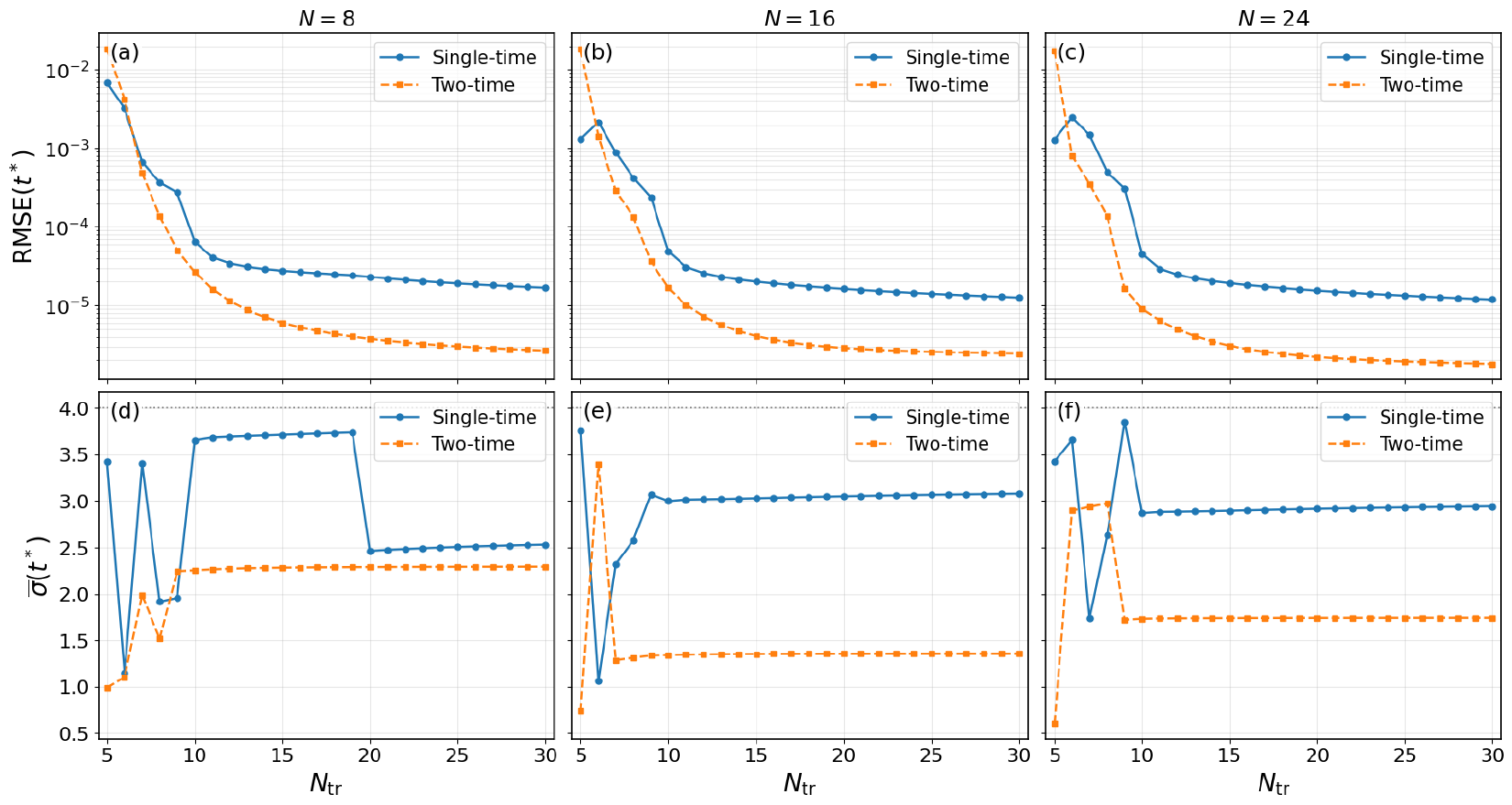}
\caption{Minimum RMSE and average uncertainty $\overline{\sigma}$ as functions of the number of training points $N_{\mathrm{tr}}$. The average uncertainty $\overline{\sigma}$ is evaluated over 200 test values of $h$ in the interval $h \in [0,2]$. The optimal evolution time $t^{*}$ is determined by minimizing the RMSE subject to the constraint $\overline{\sigma} \leq 4$. The different columns correspond to system sizes $N = 8, 16,$ and $24$, respectively. For the two-time measurement scheme, we fixed $\Delta t = 0.5/J$.}
    \label{RMSE_sigma_vstraining}
\end{figure*}
\subsection{Strategy for choosing the measurement time }

As shown before, both the RMSE and the average single-shot measurement uncertainty $\overline \sigma(t)$ are functions of time. Ideally, we would like to minimize both quantities simultaneously. However, their minima do not coincide at the same time. Thus, we need to find a proper strategy to fix the measurement time in order to achieve the best possible RMSE while keeping the sample fluctuation $\overline \sigma(t)$ small enough that does not affect the performance. 
\begin{figure*}[t]
    \includegraphics[width=0.9\textwidth]{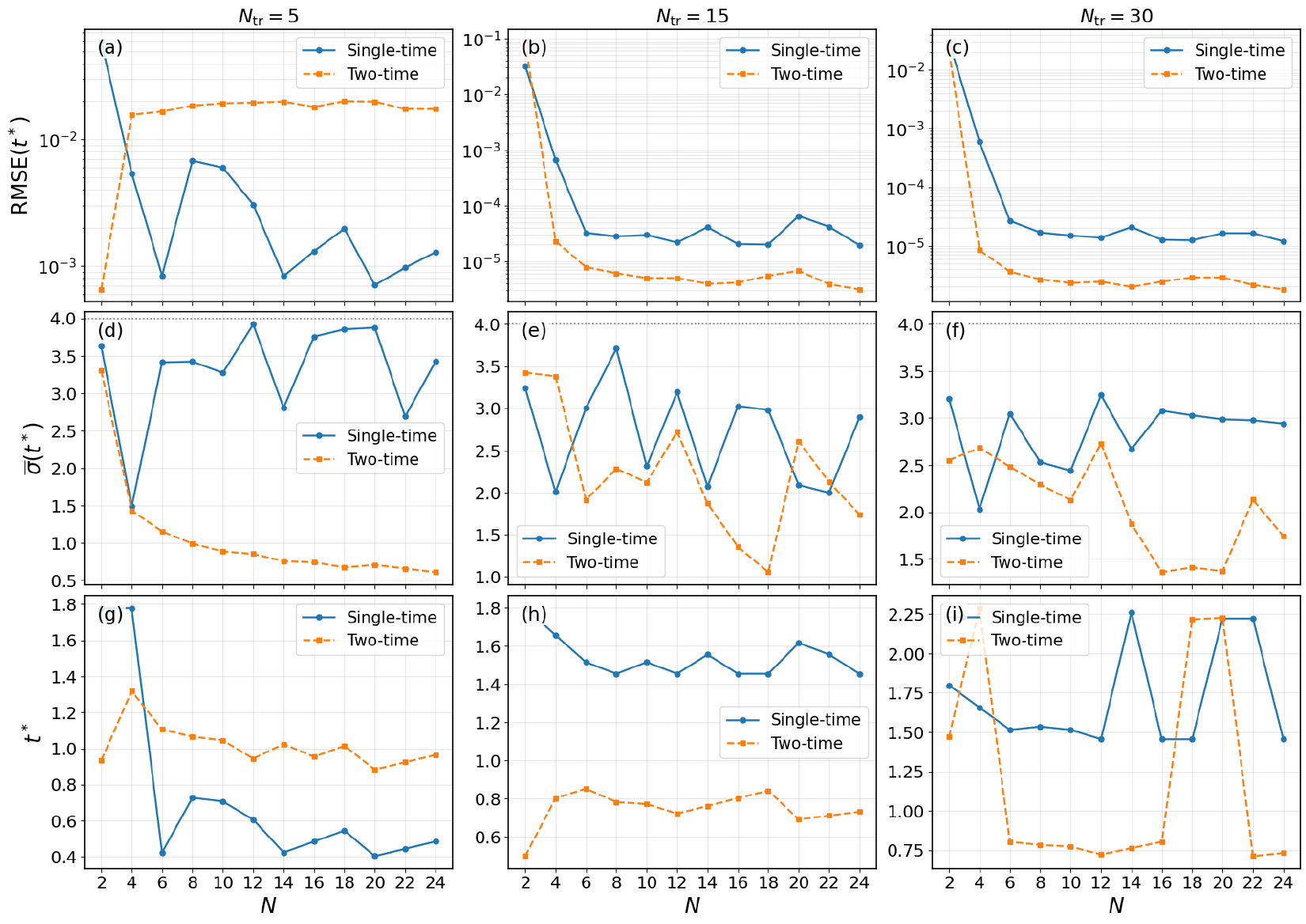}
\caption{Minimum root-mean-square error (RMSE) and average uncertainty $\overline{\sigma}$ as functions of the number of qubits $N$. The average uncertainty $\overline{\sigma}$ is evaluated over 200 test values of $h$ in the interval $h \in [0,2]$. The optimal evolution time $t^{*}$ is determined by minimizing the RMSE subject to the constraint $\overline{\sigma} \leq 4$. The different columns correspond to different numbers of training points, $N_{\mathrm{tr}} = 5, 15,$ and $30$, respectively.}
    \label{RMSE_sigma_vsqubit}
\end{figure*}
To establish a strategy for fixing the measurement time, we first impose a constraint  $\overline \sigma(t){\le} \mathcal{C}$, where $\mathcal{C}$ is a constant. This constant defines certain time intervals during which $\overline \sigma(t)$ always remains below $\mathcal{C}$. Consequently, for $M$ measurement samples, the estimation uncertainty will be lower than  $\mathcal{C}/\sqrt{M}$. 
Thus, by controlling the constant $\mathcal{C}$ and the sample size $M$, one can control the fluctuations in the final estimation. 
In our transverse Ising model, we find that $\mathcal{C}{=}4$ is a reasonable choice: it can be achieved during several time windows and simultaneously allows for a small final uncertainty while keeping the number of measurement samples at $M{\sim}10^4{-}10^5$, as typically used in many quantum experiments. Now, we minimize the RMSE (i.e. enhancing the accuracy) subject to a precision constraint ($\overline \sigma(t){\le}\mathcal{C}$) to obtain the optimal time $t^*$. This strategy results in the minimum RMSE conditioned to keeping the final uncertainty below $\mathcal{C}{/}\sqrt{M}$. Mathematically, one can summarize the strategy as
\begin{equation} \label{eq:time_optimization}
    t^*=\underset{t \in \mathcal{T} } 
{\operatorname{arg\,min}}\,\mathrm{RMSE}(t),
\end{equation}
where $\mathcal{T}$ denotes those time intervals that $\overline \sigma(t){\le} \mathcal{C}$. 

So far, we have considered a single-time measurement protocol, where the unknown parameter is estimated using the expectation values of local observables evaluated at a fixed evolution time \(t{=}t^*\). However, the information about the unknown parameter is encoded in the time evolution of the quantum system, and measurements at a single instant may not fully capture this information. This motivates us to extend the proposed scheme by incorporating measurements at two distinct evolution times. Specifically, we consider measurements at \(t-\Delta t\) and \(t+\Delta t\), where \(\Delta t\) is a fixed time interval, to keep the average time equal to $t$. The expectation values obtained at these two times are combined to construct an enlarged feature vector with $L{=}4N$, basically the double sized of the vector in Eq.~(\ref{eq:local_expectations}). Consequently, the subsequent training procedure results in a weight matrix $\mathbf{W}^{\mathrm{opt}}$ whose size is double the size of weight matrix in the case of a single-time measurement. The two-time measurement protocol is expected to be more informative about the unknown parameter $h$. This is because if the expected measurement outcomes at one time are not capable of uniquely determining the value of $h$, the measurement outcomes at the other time may compensate and enhance the performance. While we fix $\Delta t{=}0.5/J$ in the two-time measurement protocol, we use the same strategy of of Eq.~(\ref{eq:time_optimization}) for optimizing the time $t$. 
However, to have a fair comparison between one-time and two-time measurement protocols one has to keep the total number of measurements the same. Since two-time measurement protocol requires twice more operator expectations, namely $\langle \mathcal{O}_j\rangle(t)$ and $\langle \mathcal{O}_j\rangle(t+\Delta t)$, one can reduce the measurement budget for each expectation value to $M/2$. Therefore, for direct comparison between the two protocols, we renormalize the definition for  the two-time measurement protocol as 
\begin{align} 
    \sigma(h,t) &= \sqrt{2} \Bigg(
    \sum_{j=1}^{2N} \mathbf{W}_j^2
    \left[ 1-\langle\mathcal{O}_j(t)\rangle^2 \right] \nonumber \\
    &\quad + \sum_{j=2N+1}^{4N} \mathbf{W}_j^2
    \left[ 1-\langle\mathcal{O}_j(t+\Delta t)\rangle^2 \right] \Bigg)^{1/2},
\end{align}
where the $\sqrt{2}$ coefficient takes care of the measurement budgeting, explained above. One then can compute the average single-shot measurement uncertainty $\overline \sigma(t)$ for two-time measurement protocol using Eq.~(\ref{average_sigma}).

\subsection{The impact of training points}

In this part, we investigate the effect of training points on the performance of our extreme learning based sensing protocol. By choosing the optimal time $t^*$, through the procedure in Eq.~(\ref{eq:time_optimization}),  one can quantify the performance of the protocol using two main figures of merit, namely $\mathrm{RMSE}(t^*)$ and $\overline \sigma(t^*)$. In Figs.~\ref{RMSE_sigma_vstraining}(a)-(c), we plot 
$\mathrm{RMSE(t^*)}$ for both single- and two-time measurement protocols as a function of the number of training points $N_{\mathrm{tr}}$ for three different system sizes, respectively. Two clear features can be observed. First, as the number of training points $N_{\mathrm{tr}}$ increases the two-time measurement protocol achieves  significantly lower values of RMSE than the single-time measurement protocol, demonstrating the superiority of its performance. Second, both single- and two-time measurement protocols show almost monotonic improvements  by increasing the number of training points  $N_{\mathrm{tr}}$.  In Figs.~\ref{RMSE_sigma_vstraining}(d)-(f), we plot the corresponding average measurement uncertainty $\overline{\sigma}(t)$ for both single- and two-time measurement protocols  as a function of $N_{\mathrm{tr}}$ for three different system sizes, respectively. While there is no monotonic behavior, the figures clearly show that uncertainty remains bounded for all the values of $N_{\mathrm{tr}}$. This is not surprising as the time optimization protocol naturally constrains $\overline{\sigma}(t)$ to be lower than $\mathcal{C}{=}4$.

\subsection{The impact of reservoir size}
So far, we have examined the influence of the optimal evolution time and the number of training points on the estimation accuracy for a fixed reservoir size. We now turn our attention to how the system size affects the accuracy of transverse-field estimation.
In Fig.~\ref{RMSE_sigma_vsqubit}, we present the $\mathrm{RMSE}$, the corresponding average uncertainty $\overline{\sigma}$, computed at the optimal time $t^{*}$.  
As illustrated in Figs.~\ref{RMSE_sigma_vsqubit}(a)-(c), increasing $N$ generally leads to improved estimation accuracy, although this advantage becomes less pronounced as the training set grows. This behavior suggests that, beyond a certain training capacity, the additional information gained by enlarging the system becomes progressively less significant, indicating a tendency toward saturation in the estimation performance. Notably, despite the overall improvement, the RMSE does not exhibit any clear scaling behavior with $N$. A possible explanation lies in the nearest-neighbor nature of the interactions, together with the restriction to local observables as measurement features. Under these conditions, increasing the number of spins does not necessarily translate into a corresponding enhancement of the information accessible for parameter estimation. 
Consequently, increasing the system size does not necessarily provide a proportional increase in the information accessible through these measurements. 

However, the two-time protocol consistently achieves a lower RMSE than the single-time protocol for moderate to large training values, highlighting the advantage of incorporating information from multiple evolution times. The corresponding average uncertainties, shown in Figs.~\ref{RMSE_sigma_vsqubit}(d)-(f), are also generally lower for the two-time protocol, particularly for larger system sizes. Finally, Figs. ~\ref{RMSE_sigma_vsqubit}(g)-(i) illustrate the dependence of the optimal evolution time on $N$. Although the optimal times fluctuate with system size, no systematic dependence on $N$ is evident within the investigated range. 

\begin{figure*}[t]
 \centering
    \includegraphics[width=0.9\textwidth]
    {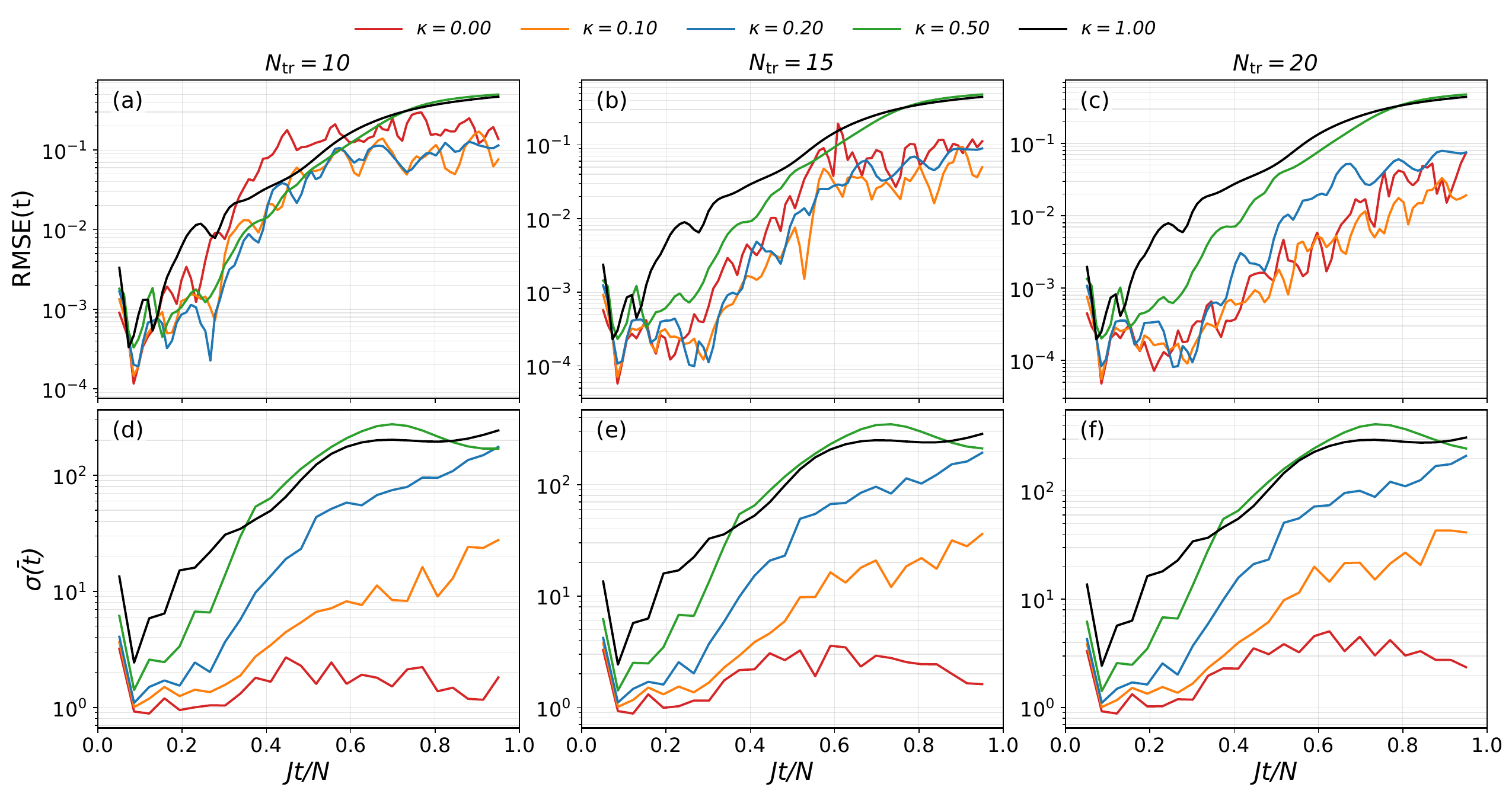}
    \caption{The effect of decoherence on the performance of RMSE as a function of normalized time $Jt/N$ in the two-time measurement protocol (with $\Delta t{=}0.5/J$) for a probe of size $N=10$ and the training point of: (a) $N_{\mathrm{tr}}{=}10$; (b) $N_{\mathrm{tr}}{=}15$; and (c) $N_{\mathrm{tr}}{=}20$. The corresponding sampling error $\overline{\sigma}(t)$ under the action of dephasing for training points of: (d) $N_{\mathrm{tr}}{=}10$; (e) $N_{\mathrm{tr}}{=}15$; and (f) $N_{\mathrm{tr}}{=}20$.}
\label{fig:decoherence}   
\end{figure*}

\section{Robustness in the presence of decoherence}
\label{decoherence}

So far, we have investigated the performance of the proposed parameter estimation protocol via QEL under unitary dynamics, where the parameter is encoded in the quantum state through coherent evolution. However, in realistic scenarios, quantum systems inevitably interact with their surrounding environment, which can affect the estimation accuracy \cite{demkowicz2012elusive, Kołodyński_2013, alipour2014, rafal2014, Ghosh2026optimalphase}. In this section, we extend our analysis to open quantum systems and investigate the effects of environmental noise on the performance of the proposed protocol. We consider local dephasing acting on all qubits while we are still interested in sensing the transverse field $h$. The evolution of the system is described by the following Lindblad master equation 
\begin{equation}
\frac{d\rho}{dt} =
-i[H(h),\rho]
+\kappa \sum_{j=1}^N
\left(Z_j\rho Z_j-\rho\right),
\end{equation}
where $\kappa$ is the strength of dephasing noise. 
In order to study the effect of dephasing, in Figs.~\ref{fig:decoherence}(a)-(c), we plot the RMSE, computed for a two-time measurement strategy, as a function of time in a reservoir of size $N{=}10$ for three different number of training points and various values of dephasing strength $\kappa$. As the figure shows, in short time scales, the effect of dephasing is negligible and the performance remains robust and only at longer time scales the dephasing becomes important and deteriorates the accuracy. In particular, for larger number of training points, the minimum of the RMSE, which is the relevant quantity for our estimation procedure, is hardly affected by dephasing, unless it is as strong as $\kappa{\simeq}J$. In Figs.~\ref{fig:decoherence}(d)-(f) we plot the corresponding single-shot sample uncertainty $\overline{\sigma}(t)$ as a function of time. A similar behavior can also be observed here, showing that dephasing increases the sample uncertainty, however, at short time scales where the sensing is performed, the protocol is quite robust and its performance is not affected much unless the noise becomes very strong such as $\kappa{\simeq}J$. This demonstrates that the QEL-based sensing protocol is quite robust against dephasing and thus can be implemented in near-term quantum devices.  

\section{conclusion}
\label{conclusion}

In this work, we have proposed a QEL-based framework for estimating unknown parameters encoded in the Hamiltonian of a many-body quantum system. Unlike conventional QEL protocols, which encode input information in the initial quantum state, our approach employs a fixed input state and encodes the unknown parameter through the subsequent system dynamics. By combining the expectation values of local observables measured after the reservoir evolution with a trainable classical readout, the proposed framework enables parameter estimation without requiring the explicit construction or evaluation of computationally demanding likelihood functions. 

To demonstrate the effectiveness of the proposed approach, we have considered the transverse-field Ising model, where the transverse magnetic field is treated as the unknown parameter. We have four main results. First, our numerical results show that accurate parameter estimation can be achieved using a relatively small number of training points. Second, we have developed an interrogation-time optimization method which  minimizes the inaccuracy, quantified by RMSE, while ensuring that sample fluctuations remain below a prescribed threshold. This optimization show that the protocol is very efficient achieving high accuracy within a fairly short time scale. Third,  by extending our framework to a two-time measurement protocol, the accuracy is significantly improved while the sampling uncertainty still remains below the prescribed threshold.  Fourth, our protocol is shown to be robust against  dephasing making it suitable for realizing on near-term quantum devices. Our results, show that QEL can provide an effective alternative approach to optimal estimator protocols, such as Bayesian inference and maximum likelihood functions, which rely on construction of resource demanding  likelihood functions.


\section{Acknowledgment}
A.B. acknowledges support from the National Natural Science Foundation of China (grants No.~W2541020, No.~12274059, No.~12574528, 
and No.~1251101297). V.M. thanks support from the National Natural Science Foundation of China Grants No. 12374482 and No. W2432005.

\appendix

\section{Diagonalization of the Hamiltonian and evaluation of observables}
\label{diagonalization}

Here, we describe the procedure used to
evaluate the local spin expectation values entering the quantum
reservoir feature vector. Although the transverse-field Ising
model is exactly solvable by mapping it onto a quadratic
fermionic Hamiltonian, the evaluation of local observables
requires additional care because of the particular initial
state considered in this work. In particular, the expectation
values of $\sigma_j^x$ involve odd products of Majorana
operators, which cannot be evaluated using the conventional
covariance-matrix formulation of Wick's theorem alone.

\subsection*{Mapping the transverse-field Ising Hamiltonian
to Majorana fermions}

We consider the transverse-field Ising Hamiltonian with open
boundary conditions,
\begin{equation}
    H =  - J \sum_{i = 1}^{N-1} {X}_{i} X_{i+1}- h \sum_{i = 1}^{N} {Z}_{i},
    \label{appendix:tfim}
\end{equation}
where $J_{\mathrm{int}}$ is the nearest-neighbor interaction strength and
$h_{\mathrm{loc}}$ is the transverse magnetic field to be estimated.

To express the Hamiltonian in terms of fermionic degrees
of freedom, we introduce the Jordan--Wigner Majorana operators
\begin{align}
\gamma_{2j-1}
&=
\left(\prod_{k=1}^{j-1}\sigma_k^z\right)
\sigma_j^x,
\\
\gamma_{2j}
&=
\left(\prod_{k=1}^{j-1}\sigma_k^z\right)
\sigma_j^y,
\end{align}
for $j=1,\ldots,N$. These operators are Hermitian and obey
the canonical anticommutation relations
\begin{equation}
\{\gamma_a,\gamma_b\}=2\delta_{ab}.
\end{equation}

Using the Pauli algebra, one obtains
\begin{align}
\sigma_j^z
&=-i\gamma_{2j-1}\gamma_{2j},
\\
\sigma_j^x\sigma_{j+1}^x
&=-i\gamma_{2j}\gamma_{2j+1}.
\end{align}
Substituting these identities into
Eq.~\eqref{appendix:tfim}, the Hamiltonian becomes
\begin{equation}
H=
iJ\sum_{j=1}^{N-1}\gamma_{2j}\gamma_{2j+1}
+ih\sum_{j=1}^{N}\gamma_{2j-1}\gamma_{2j}.
\label{eq:majorana_hamiltonian}
\end{equation}

Introducing the Majorana vector
\begin{equation}
\boldsymbol{\gamma}
=
(\gamma_1,\gamma_2,\ldots,\gamma_{2N})^T,
\end{equation}
we can write the Hamiltonian in the compact quadratic form
\begin{equation}
H=\frac{i}{4}
\boldsymbol{\gamma}^{T}A\boldsymbol{\gamma},
\label{eq:quadratic_majorana}
\end{equation}
where $A$ is a real antisymmetric matrix of dimension
$2N\times 2N$.

The nonzero entries of $A$ are
\begin{align}
A_{2j-1,2j}&=2h,
&&j=1,\ldots,N,
\\
A_{2j,2j+1}&=2J,
&&j=1,\ldots,N-1,
\end{align}
together with $A_{ba}=-A_{ab}$.

Thus, the interacting spin Hamiltonian is transformed into
a quadratic fermionic Hamiltonian. This transformation is
particularly useful because the dynamics can be determined
from a matrix whose dimension grows linearly with the
number of spins, rather than from a Hilbert-space matrix
of dimension $2^N$.

\subsection{Time evolution of Majorana operators}

For a quadratic Hamiltonian of the form
Eq.~\eqref{eq:quadratic_majorana}, the Heisenberg equations
of motion are linear:
\begin{equation}
\frac{d\boldsymbol{\gamma}(t)}{dt}
=A\boldsymbol{\gamma}(t).
\end{equation}
Consequently,
\begin{equation}
\boldsymbol{\gamma}(t)
=R(t)\boldsymbol{\gamma}(0),
\qquad
R(t)=e^{At}.
\label{eq:majorana_evolution}
\end{equation}

Since $A$ is real and antisymmetric, $R(t)$ is a real
orthogonal matrix.

For numerical efficiency, we diagonalize the Hermitian
matrix $iA$,
\begin{equation}
iA=U\Lambda U^\dagger,
\end{equation}
where $\Lambda$ is a real diagonal matrix. The rotation
matrix is then evaluated as
\begin{equation}
R(t)=Ue^{-i\Lambda t}U^\dagger.
\label{eq:majorana_rotation}
\end{equation}
Although the right-hand side is expressed using complex
matrices, the resulting rotation matrix is real up to
numerical roundoff.

Importantly, this procedure allows the same eigendecomposition
to be reused for different evolution times at fixed values
of $h$ and $J$.



\bibliographystyle{apsrev4-2}
\bibliography{bibliography}

\begin{thebibliography}{67}%
\makeatletter
\providecommand \@ifxundefined [1]{%
 \@ifx{#1\undefined}
}%
\providecommand \@ifnum [1]{%
 \ifnum #1\expandafter \@firstoftwo
 \else \expandafter \@secondoftwo
 \fi
}%
\providecommand \@ifx [1]{%
 \ifx #1\expandafter \@firstoftwo
 \else \expandafter \@secondoftwo
 \fi
}%
\providecommand \natexlab [1]{#1}%
\providecommand \enquote  [1]{``#1''}%
\providecommand \bibnamefont  [1]{#1}%
\providecommand \bibfnamefont [1]{#1}%
\providecommand \citenamefont [1]{#1}%
\providecommand \href@noop [0]{\@secondoftwo}%
\providecommand \href [0]{\begingroup \@sanitize@url \@href}%
\providecommand \@href[1]{\@@startlink{#1}\@@href}%
\providecommand \@@href[1]{\endgroup#1\@@endlink}%
\providecommand \@sanitize@url [0]{\catcode `\\12\catcode `\$12\catcode `\&12\catcode `\#12\catcode `\^12\catcode `\_12\catcode `\%12\relax}%
\providecommand \@@startlink[1]{}%
\providecommand \@@endlink[0]{}%
\providecommand \url  [0]{\begingroup\@sanitize@url \@url }%
\providecommand \@url [1]{\endgroup\@href {#1}{\urlprefix }}%
\providecommand \urlprefix  [0]{URL }%
\providecommand \Eprint [0]{\href }%
\providecommand \doibase [0]{https://doi.org/}%
\providecommand \selectlanguage [0]{\@gobble}%
\providecommand \bibinfo  [0]{\@secondoftwo}%
\providecommand \bibfield  [0]{\@secondoftwo}%
\providecommand \translation [1]{[#1]}%
\providecommand \BibitemOpen [0]{}%
\providecommand \bibitemStop [0]{}%
\providecommand \bibitemNoStop [0]{.\EOS\space}%
\providecommand \EOS [0]{\spacefactor3000\relax}%
\providecommand \BibitemShut  [1]{\csname bibitem#1\endcsname}%
\let\auto@bib@innerbib\@empty
\bibitem [{\citenamefont {Degen}\ \emph {et~al.}(2017)\citenamefont {Degen}, \citenamefont {Reinhard},\ and\ \citenamefont {Cappellaro}}]{degen2017quantum}%
  \BibitemOpen
  \bibfield  {author} {\bibinfo {author} {\bibfnamefont {C.~L.}\ \bibnamefont {Degen}}, \bibinfo {author} {\bibfnamefont {F.}~\bibnamefont {Reinhard}},\ and\ \bibinfo {author} {\bibfnamefont {P.}~\bibnamefont {Cappellaro}},\ }\href {https://doi.org/10.1103/RevModPhys.89.035002} {\bibfield  {journal} {\bibinfo  {journal} {Rev. Mod. Phys.}\ }\textbf {\bibinfo {volume} {89}},\ \bibinfo {pages} {035002} (\bibinfo {year} {2017})}\BibitemShut {NoStop}%
\bibitem [{\citenamefont {Pezz{\`e}}\ \emph {et~al.}(2018)\citenamefont {Pezz{\`e}}, \citenamefont {Smerzi}, \citenamefont {Oberthaler}, \citenamefont {Schmied},\ and\ \citenamefont {Treutlein}}]{pezze2018quantum}%
  \BibitemOpen
  \bibfield  {author} {\bibinfo {author} {\bibfnamefont {L.}~\bibnamefont {Pezz{\`e}}}, \bibinfo {author} {\bibfnamefont {A.}~\bibnamefont {Smerzi}}, \bibinfo {author} {\bibfnamefont {M.~K.}\ \bibnamefont {Oberthaler}}, \bibinfo {author} {\bibfnamefont {R.}~\bibnamefont {Schmied}},\ and\ \bibinfo {author} {\bibfnamefont {P.}~\bibnamefont {Treutlein}},\ }\href {https://doi.org/10.1103/RevModPhys.90.035005} {\bibfield  {journal} {\bibinfo  {journal} {Rev. Mod. Phys.}\ }\textbf {\bibinfo {volume} {90}},\ \bibinfo {pages} {035005} (\bibinfo {year} {2018})}\BibitemShut {NoStop}%
\bibitem [{\citenamefont {Braun}\ \emph {et~al.}(2018)\citenamefont {Braun}, \citenamefont {Adesso}, \citenamefont {Benatti}, \citenamefont {Floreanini}, \citenamefont {Marzolino}, \citenamefont {Mitchell},\ and\ \citenamefont {Pirandola}}]{braun2018quantum}%
  \BibitemOpen
  \bibfield  {author} {\bibinfo {author} {\bibfnamefont {D.}~\bibnamefont {Braun}}, \bibinfo {author} {\bibfnamefont {G.}~\bibnamefont {Adesso}}, \bibinfo {author} {\bibfnamefont {F.}~\bibnamefont {Benatti}}, \bibinfo {author} {\bibfnamefont {R.}~\bibnamefont {Floreanini}}, \bibinfo {author} {\bibfnamefont {U.}~\bibnamefont {Marzolino}}, \bibinfo {author} {\bibfnamefont {M.~W.}\ \bibnamefont {Mitchell}},\ and\ \bibinfo {author} {\bibfnamefont {S.}~\bibnamefont {Pirandola}},\ }\href {https://doi.org/10.1103/RevModPhys.90.035006} {\bibfield  {journal} {\bibinfo  {journal} {Rev. Mod. Phys.}\ }\textbf {\bibinfo {volume} {90}},\ \bibinfo {pages} {035006} (\bibinfo {year} {2018})}\BibitemShut {NoStop}%
\bibitem [{\citenamefont {Ye}\ and\ \citenamefont {Zoller}(2024)}]{ye2024essay}%
  \BibitemOpen
  \bibfield  {author} {\bibinfo {author} {\bibfnamefont {J.}~\bibnamefont {Ye}}\ and\ \bibinfo {author} {\bibfnamefont {P.}~\bibnamefont {Zoller}},\ }\href {https://doi.org/10.1103/PhysRevLett.132.190001} {\bibfield  {journal} {\bibinfo  {journal} {Phys. Rev. Lett.}\ }\textbf {\bibinfo {volume} {132}},\ \bibinfo {pages} {190001} (\bibinfo {year} {2024})}\BibitemShut {NoStop}%
\bibitem [{\citenamefont {Ghosh}\ \emph {et~al.}(2026{\natexlab{a}})\citenamefont {Ghosh}, \citenamefont {Konar}, \citenamefont {Rakshit}, \citenamefont {De},\ and\ \citenamefont {Sen}}]{ghosh2026journey}%
  \BibitemOpen
  \bibfield  {author} {\bibinfo {author} {\bibfnamefont {P.}~\bibnamefont {Ghosh}}, \bibinfo {author} {\bibfnamefont {T.~K.}\ \bibnamefont {Konar}}, \bibinfo {author} {\bibfnamefont {D.}~\bibnamefont {Rakshit}}, \bibinfo {author} {\bibfnamefont {A.~S.}\ \bibnamefont {De}},\ and\ \bibinfo {author} {\bibfnamefont {U.}~\bibnamefont {Sen}},\ }\href {https://arxiv.org/abs/2605.21702} {\bibinfo {title} {Journey in quantum metrology and sensing from foundations to applications: a review}} (\bibinfo {year} {2026}{\natexlab{a}}),\ \Eprint {https://arxiv.org/abs/2605.21702} {arXiv:2605.21702 [quant-ph]} \BibitemShut {NoStop}%
\bibitem [{\citenamefont {Giovannetti}\ \emph {et~al.}(2004)\citenamefont {Giovannetti}, \citenamefont {Lloyd},\ and\ \citenamefont {Maccone}}]{giovannetti2004quantum}%
  \BibitemOpen
  \bibfield  {author} {\bibinfo {author} {\bibfnamefont {V.}~\bibnamefont {Giovannetti}}, \bibinfo {author} {\bibfnamefont {S.}~\bibnamefont {Lloyd}},\ and\ \bibinfo {author} {\bibfnamefont {L.}~\bibnamefont {Maccone}},\ }\href {https://doi.org/10.1126/science.1104149} {\bibfield  {journal} {\bibinfo  {journal} {Science}\ }\textbf {\bibinfo {volume} {306}},\ \bibinfo {pages} {1330} (\bibinfo {year} {2004})}\BibitemShut {NoStop}%
\bibitem [{\citenamefont {Giovannetti}\ \emph {et~al.}(2011)\citenamefont {Giovannetti}, \citenamefont {Lloyd},\ and\ \citenamefont {Maccone}}]{giovannetti2011advances}%
  \BibitemOpen
  \bibfield  {author} {\bibinfo {author} {\bibfnamefont {V.}~\bibnamefont {Giovannetti}}, \bibinfo {author} {\bibfnamefont {S.}~\bibnamefont {Lloyd}},\ and\ \bibinfo {author} {\bibfnamefont {L.}~\bibnamefont {Maccone}},\ }\href {https://doi.org/10.1038/nphoton.2011.35} {\bibfield  {journal} {\bibinfo  {journal} {Nature photonics}\ }\textbf {\bibinfo {volume} {5}},\ \bibinfo {pages} {222} (\bibinfo {year} {2011})}\BibitemShut {NoStop}%
\bibitem [{\citenamefont {Giovannetti}\ \emph {et~al.}(2006)\citenamefont {Giovannetti}, \citenamefont {Lloyd},\ and\ \citenamefont {Maccone}}]{giovannetti2006quantum}%
  \BibitemOpen
  \bibfield  {author} {\bibinfo {author} {\bibfnamefont {V.}~\bibnamefont {Giovannetti}}, \bibinfo {author} {\bibfnamefont {S.}~\bibnamefont {Lloyd}},\ and\ \bibinfo {author} {\bibfnamefont {L.}~\bibnamefont {Maccone}},\ }\href {https://doi.org/10.1103/PhysRevLett.96.010401} {\bibfield  {journal} {\bibinfo  {journal} {Phys. Rev. Lett.}\ }\textbf {\bibinfo {volume} {96}},\ \bibinfo {pages} {010401} (\bibinfo {year} {2006})}\BibitemShut {NoStop}%
\bibitem [{\citenamefont {Pezz{\'e}}\ and\ \citenamefont {Smerzi}(2008)}]{pezze2008mach}%
  \BibitemOpen
  \bibfield  {author} {\bibinfo {author} {\bibfnamefont {L.}~\bibnamefont {Pezz{\'e}}}\ and\ \bibinfo {author} {\bibfnamefont {A.}~\bibnamefont {Smerzi}},\ }\href {https://doi.org/10.1103/PhysRevLett.100.073601} {\bibfield  {journal} {\bibinfo  {journal} {Phys. Rev. Lett.}\ }\textbf {\bibinfo {volume} {100}},\ \bibinfo {pages} {073601} (\bibinfo {year} {2008})}\BibitemShut {NoStop}%
\bibitem [{\citenamefont {Montenegro}\ \emph {et~al.}(2025)\citenamefont {Montenegro}, \citenamefont {Mukhopadhyay}, \citenamefont {Yousefjani}, \citenamefont {Sarkar}, \citenamefont {Mishra}, \citenamefont {Paris},\ and\ \citenamefont {Bayat}}]{montenegro2025quantum}%
  \BibitemOpen
  \bibfield  {author} {\bibinfo {author} {\bibfnamefont {V.}~\bibnamefont {Montenegro}}, \bibinfo {author} {\bibfnamefont {C.}~\bibnamefont {Mukhopadhyay}}, \bibinfo {author} {\bibfnamefont {R.}~\bibnamefont {Yousefjani}}, \bibinfo {author} {\bibfnamefont {S.}~\bibnamefont {Sarkar}}, \bibinfo {author} {\bibfnamefont {U.}~\bibnamefont {Mishra}}, \bibinfo {author} {\bibfnamefont {M.~G.}\ \bibnamefont {Paris}},\ and\ \bibinfo {author} {\bibfnamefont {A.}~\bibnamefont {Bayat}},\ }\href {https://doi.org/https://doi.org/10.1016/j.physrep.2025.05.005} {\bibfield  {journal} {\bibinfo  {journal} {Physics Reports}\ }\textbf {\bibinfo {volume} {1134}},\ \bibinfo {pages} {1} (\bibinfo {year} {2025})}\BibitemShut {NoStop}%
\bibitem [{\citenamefont {Mihailescu}\ \emph {et~al.}(2026)\citenamefont {Mihailescu}, \citenamefont {Alushi}, \citenamefont {Di~Candia}, \citenamefont {Felicetti},\ and\ \citenamefont {Gietka}}]{mihailescu2026critical}%
  \BibitemOpen
  \bibfield  {author} {\bibinfo {author} {\bibfnamefont {G.}~\bibnamefont {Mihailescu}}, \bibinfo {author} {\bibfnamefont {U.}~\bibnamefont {Alushi}}, \bibinfo {author} {\bibfnamefont {R.}~\bibnamefont {Di~Candia}}, \bibinfo {author} {\bibfnamefont {S.}~\bibnamefont {Felicetti}},\ and\ \bibinfo {author} {\bibfnamefont {K.}~\bibnamefont {Gietka}},\ }\href {https://doi.org/10.1103/v7mf-yh8n} {\bibfield  {journal} {\bibinfo  {journal} {PRX Quantum}\ }\textbf {\bibinfo {volume} {7}},\ \bibinfo {pages} {020201} (\bibinfo {year} {2026})}\BibitemShut {NoStop}%
\bibitem [{\citenamefont {Hradil}\ \emph {et~al.}(1996)\citenamefont {Hradil}, \citenamefont {My{\ifmmode \check{s}\else \v{s}\fi{}}ka}, \citenamefont {Pe{\ifmmode \check{r}\else \v{r}\fi{}}ina}, \citenamefont {Zawisky}, \citenamefont {Hasegawa},\ and\ \citenamefont {Rauch}}]{hradil1996quantum}%
  \BibitemOpen
  \bibfield  {author} {\bibinfo {author} {\bibfnamefont {Z.}~\bibnamefont {Hradil}}, \bibinfo {author} {\bibfnamefont {R.}~\bibnamefont {My{\ifmmode \check{s}\else \v{s}\fi{}}ka}}, \bibinfo {author} {\bibfnamefont {J.}~\bibnamefont {Pe{\ifmmode \check{r}\else \v{r}\fi{}}ina}}, \bibinfo {author} {\bibfnamefont {M.}~\bibnamefont {Zawisky}}, \bibinfo {author} {\bibfnamefont {Y.}~\bibnamefont {Hasegawa}},\ and\ \bibinfo {author} {\bibfnamefont {H.}~\bibnamefont {Rauch}},\ }\href {https://doi.org/10.1103/PhysRevLett.76.4295} {\bibfield  {journal} {\bibinfo  {journal} {Phys. Rev. Lett.}\ }\textbf {\bibinfo {volume} {76}},\ \bibinfo {pages} {4295} (\bibinfo {year} {1996})}\BibitemShut {NoStop}%
\bibitem [{\citenamefont {Pezz{\'e}}\ \emph {et~al.}(2007)\citenamefont {Pezz{\'e}}, \citenamefont {Smerzi}, \citenamefont {Khoury}, \citenamefont {Hodelin},\ and\ \citenamefont {Bouwmeester}}]{pezze2007phase}%
  \BibitemOpen
  \bibfield  {author} {\bibinfo {author} {\bibfnamefont {L.}~\bibnamefont {Pezz{\'e}}}, \bibinfo {author} {\bibfnamefont {A.}~\bibnamefont {Smerzi}}, \bibinfo {author} {\bibfnamefont {G.}~\bibnamefont {Khoury}}, \bibinfo {author} {\bibfnamefont {J.~F.}\ \bibnamefont {Hodelin}},\ and\ \bibinfo {author} {\bibfnamefont {D.}~\bibnamefont {Bouwmeester}},\ }\href {https://doi.org/10.1103/PhysRevLett.99.223602} {\bibfield  {journal} {\bibinfo  {journal} {Phys. Rev. Lett.}\ }\textbf {\bibinfo {volume} {99}},\ \bibinfo {pages} {223602} (\bibinfo {year} {2007})}\BibitemShut {NoStop}%
\bibitem [{\citenamefont {Olivares}\ and\ \citenamefont {Paris}(2009)}]{olivares2009bayesian}%
  \BibitemOpen
  \bibfield  {author} {\bibinfo {author} {\bibfnamefont {S.}~\bibnamefont {Olivares}}\ and\ \bibinfo {author} {\bibfnamefont {M.~G.~A.}\ \bibnamefont {Paris}},\ }\href {https://doi.org/10.1088/0953-4075/42/5/055506} {\bibfield  {journal} {\bibinfo  {journal} {Journal of Physics B: Atomic, Molecular and Optical Physics}\ }\textbf {\bibinfo {volume} {42}},\ \bibinfo {pages} {055506} (\bibinfo {year} {2009})}\BibitemShut {NoStop}%
\bibitem [{\citenamefont {Brivio}\ \emph {et~al.}(2010)\citenamefont {Brivio}, \citenamefont {Cialdi}, \citenamefont {Vezzoli}, \citenamefont {Gebrehiwot}, \citenamefont {Genoni}, \citenamefont {Olivares},\ and\ \citenamefont {Paris}}]{Brivio2010}%
  \BibitemOpen
  \bibfield  {author} {\bibinfo {author} {\bibfnamefont {D.}~\bibnamefont {Brivio}}, \bibinfo {author} {\bibfnamefont {S.}~\bibnamefont {Cialdi}}, \bibinfo {author} {\bibfnamefont {S.}~\bibnamefont {Vezzoli}}, \bibinfo {author} {\bibfnamefont {B.~T.}\ \bibnamefont {Gebrehiwot}}, \bibinfo {author} {\bibfnamefont {M.~G.}\ \bibnamefont {Genoni}}, \bibinfo {author} {\bibfnamefont {S.}~\bibnamefont {Olivares}},\ and\ \bibinfo {author} {\bibfnamefont {M.~G.~A.}\ \bibnamefont {Paris}},\ }\href {https://doi.org/10.1103/PhysRevA.81.012305} {\bibfield  {journal} {\bibinfo  {journal} {Phys. Rev. A}\ }\textbf {\bibinfo {volume} {81}},\ \bibinfo {pages} {012305} (\bibinfo {year} {2010})}\BibitemShut {NoStop}%
\bibitem [{\citenamefont {Genoni}\ \emph {et~al.}(2012)\citenamefont {Genoni}, \citenamefont {Olivares}, \citenamefont {Brivio}, \citenamefont {Cialdi}, \citenamefont {Cipriani}, \citenamefont {Santamato}, \citenamefont {Vezzoli},\ and\ \citenamefont {Paris}}]{Genoni2012}%
  \BibitemOpen
  \bibfield  {author} {\bibinfo {author} {\bibfnamefont {M.~G.}\ \bibnamefont {Genoni}}, \bibinfo {author} {\bibfnamefont {S.}~\bibnamefont {Olivares}}, \bibinfo {author} {\bibfnamefont {D.}~\bibnamefont {Brivio}}, \bibinfo {author} {\bibfnamefont {S.}~\bibnamefont {Cialdi}}, \bibinfo {author} {\bibfnamefont {D.}~\bibnamefont {Cipriani}}, \bibinfo {author} {\bibfnamefont {A.}~\bibnamefont {Santamato}}, \bibinfo {author} {\bibfnamefont {S.}~\bibnamefont {Vezzoli}},\ and\ \bibinfo {author} {\bibfnamefont {M.~G.~A.}\ \bibnamefont {Paris}},\ }\href {https://doi.org/10.1103/PhysRevA.85.043817} {\bibfield  {journal} {\bibinfo  {journal} {Phys. Rev. A}\ }\textbf {\bibinfo {volume} {85}},\ \bibinfo {pages} {043817} (\bibinfo {year} {2012})}\BibitemShut {NoStop}%
\bibitem [{\citenamefont {Blandino}\ \emph {et~al.}(2012)\citenamefont {Blandino}, \citenamefont {Genoni}, \citenamefont {Etesse}, \citenamefont {Barbieri}, \citenamefont {Paris}, \citenamefont {Grangier},\ and\ \citenamefont {Tualle-Brouri}}]{Blandino2012}%
  \BibitemOpen
  \bibfield  {author} {\bibinfo {author} {\bibfnamefont {R.}~\bibnamefont {Blandino}}, \bibinfo {author} {\bibfnamefont {M.~G.}\ \bibnamefont {Genoni}}, \bibinfo {author} {\bibfnamefont {J.}~\bibnamefont {Etesse}}, \bibinfo {author} {\bibfnamefont {M.}~\bibnamefont {Barbieri}}, \bibinfo {author} {\bibfnamefont {M.~G.~A.}\ \bibnamefont {Paris}}, \bibinfo {author} {\bibfnamefont {P.}~\bibnamefont {Grangier}},\ and\ \bibinfo {author} {\bibfnamefont {R.}~\bibnamefont {Tualle-Brouri}},\ }\href {https://doi.org/10.1103/PhysRevLett.109.180402} {\bibfield  {journal} {\bibinfo  {journal} {Phys. Rev. Lett.}\ }\textbf {\bibinfo {volume} {109}},\ \bibinfo {pages} {180402} (\bibinfo {year} {2012})}\BibitemShut {NoStop}%
\bibitem [{\citenamefont {Le~Cam}(2012)}]{le2012asymptotic}%
  \BibitemOpen
  \bibfield  {author} {\bibinfo {author} {\bibfnamefont {L.}~\bibnamefont {Le~Cam}},\ }\href {https://doi.org/https://doi.org/10.1007/978-1-4612-4946-7} {\emph {\bibinfo {title} {Asymptotic methods in statistical decision theory}}}\ (\bibinfo  {publisher} {Springer},\ \bibinfo {address} {New York},\ \bibinfo {year} {2012})\BibitemShut {NoStop}%
\bibitem [{\citenamefont {Rubio}\ and\ \citenamefont {Dunningham}(2019)}]{rubio2019quantum}%
  \BibitemOpen
  \bibfield  {author} {\bibinfo {author} {\bibfnamefont {J.}~\bibnamefont {Rubio}}\ and\ \bibinfo {author} {\bibfnamefont {J.}~\bibnamefont {Dunningham}},\ }\href {https://doi.org/10.1088/1367-2630/ab098b} {\bibfield  {journal} {\bibinfo  {journal} {New Journal of Physics}\ }\textbf {\bibinfo {volume} {21}},\ \bibinfo {pages} {043037} (\bibinfo {year} {2019})}\BibitemShut {NoStop}%
\bibitem [{\citenamefont {Gianani}\ \emph {et~al.}(2020)\citenamefont {Gianani}, \citenamefont {Genoni},\ and\ \citenamefont {Barbieri}}]{Gianani2020}%
  \BibitemOpen
  \bibfield  {author} {\bibinfo {author} {\bibfnamefont {I.}~\bibnamefont {Gianani}}, \bibinfo {author} {\bibfnamefont {M.~G.}\ \bibnamefont {Genoni}},\ and\ \bibinfo {author} {\bibfnamefont {M.}~\bibnamefont {Barbieri}},\ }\href {https://doi.org/10.1109/JSTQE.2020.2982976} {\bibfield  {journal} {\bibinfo  {journal} {IEEE Journal of Selected Topics in Quantum Electronics}\ }\textbf {\bibinfo {volume} {26}},\ \bibinfo {pages} {1} (\bibinfo {year} {2020})}\BibitemShut {NoStop}%
\bibitem [{\citenamefont {Yu}\ \emph {et~al.}(2025)\citenamefont {Yu}, \citenamefont {Liu}, \citenamefont {Xue}, \citenamefont {Yang}, \citenamefont {Wang}, \citenamefont {Zhang}, \citenamefont {Cui}, \citenamefont {Yang}, \citenamefont {Li}, \citenamefont {Han},\ and\ \citenamefont {Yu}}]{yu2025experimental}%
  \BibitemOpen
  \bibfield  {author} {\bibinfo {author} {\bibfnamefont {Y.}~\bibnamefont {Yu}}, \bibinfo {author} {\bibfnamefont {R.}~\bibnamefont {Liu}}, \bibinfo {author} {\bibfnamefont {G.}~\bibnamefont {Xue}}, \bibinfo {author} {\bibfnamefont {C.}~\bibnamefont {Yang}}, \bibinfo {author} {\bibfnamefont {C.}~\bibnamefont {Wang}}, \bibinfo {author} {\bibfnamefont {J.}~\bibnamefont {Zhang}}, \bibinfo {author} {\bibfnamefont {J.}~\bibnamefont {Cui}}, \bibinfo {author} {\bibfnamefont {X.}~\bibnamefont {Yang}}, \bibinfo {author} {\bibfnamefont {J.}~\bibnamefont {Li}}, \bibinfo {author} {\bibfnamefont {J.}~\bibnamefont {Han}},\ and\ \bibinfo {author} {\bibfnamefont {H.}~\bibnamefont {Yu}},\ }\href {https://arxiv.org/abs/2501.04955} {\bibinfo {title} {Experimental realization of criticality-enhanced global quantum sensing via non-equilibrium dynamics}} (\bibinfo {year} {2025}),\ \Eprint {https://arxiv.org/abs/2501.04955} {arXiv:2501.04955 [quant-ph]} \BibitemShut {NoStop}%
\bibitem [{\citenamefont {Li}\ \emph {et~al.}(2026{\natexlab{a}})\citenamefont {Li}, \citenamefont {Yang}, \citenamefont {Shi}, \citenamefont {Wang}, \citenamefont {Wang}, \citenamefont {Li}, \citenamefont {Zhang}, \citenamefont {Zhao}, \citenamefont {Xu}, \citenamefont {Deng}, \citenamefont {Liu}, \citenamefont {Ma}, \citenamefont {Li}, \citenamefont {Zhang}, \citenamefont {Fang}, \citenamefont {Song}, \citenamefont {Liu}, \citenamefont {Zhou}, \citenamefont {Liu}, \citenamefont {Chen}, \citenamefont {Liang}, \citenamefont {Song}, \citenamefont {Xiang}, \citenamefont {Xu}, \citenamefont {Huang}, \citenamefont {Bayat},\ and\ \citenamefont {Fan}}]{li2026non}%
  \BibitemOpen
  \bibfield  {author} {\bibinfo {author} {\bibfnamefont {H.}~\bibnamefont {Li}}, \bibinfo {author} {\bibfnamefont {Y.}~\bibnamefont {Yang}}, \bibinfo {author} {\bibfnamefont {Y.-H.}\ \bibnamefont {Shi}}, \bibinfo {author} {\bibfnamefont {Z.-A.}\ \bibnamefont {Wang}}, \bibinfo {author} {\bibfnamefont {Z.}~\bibnamefont {Wang}}, \bibinfo {author} {\bibfnamefont {J.}~\bibnamefont {Li}}, \bibinfo {author} {\bibfnamefont {Y.}~\bibnamefont {Zhang}}, \bibinfo {author} {\bibfnamefont {K.}~\bibnamefont {Zhao}}, \bibinfo {author} {\bibfnamefont {Y.}~\bibnamefont {Xu}}, \bibinfo {author} {\bibfnamefont {C.-L.}\ \bibnamefont {Deng}}, \bibinfo {author} {\bibfnamefont {Y.}~\bibnamefont {Liu}}, \bibinfo {author} {\bibfnamefont {W.-G.}\ \bibnamefont {Ma}}, \bibinfo {author} {\bibfnamefont {T.-M.}\ \bibnamefont {Li}}, \bibinfo {author} {\bibfnamefont {J.}~\bibnamefont {Zhang}}, \bibinfo {author} {\bibfnamefont {C.-P.}\ \bibnamefont {Fang}}, \bibinfo {author} {\bibfnamefont {J.-C.}\ \bibnamefont {Song}}, \bibinfo {author}
  {\bibfnamefont {H.-T.}\ \bibnamefont {Liu}}, \bibinfo {author} {\bibfnamefont {S.-Y.}\ \bibnamefont {Zhou}}, \bibinfo {author} {\bibfnamefont {Z.-H.}\ \bibnamefont {Liu}}, \bibinfo {author} {\bibfnamefont {B.-J.}\ \bibnamefont {Chen}}, \bibinfo {author} {\bibfnamefont {G.-H.}\ \bibnamefont {Liang}}, \bibinfo {author} {\bibfnamefont {X.}~\bibnamefont {Song}}, \bibinfo {author} {\bibfnamefont {Z.}~\bibnamefont {Xiang}}, \bibinfo {author} {\bibfnamefont {K.}~\bibnamefont {Xu}}, \bibinfo {author} {\bibfnamefont {K.}~\bibnamefont {Huang}}, \bibinfo {author} {\bibfnamefont {A.}~\bibnamefont {Bayat}},\ and\ \bibinfo {author} {\bibfnamefont {H.}~\bibnamefont {Fan}},\ }\href {https://doi.org/https://doi.org/10.1016/j.scib.2026.05.023} {\bibfield  {journal} {\bibinfo  {journal} {Science Bulletin}\ }\textbf {\bibinfo {volume} {71}},\ \bibinfo {pages} {2996} (\bibinfo {year} {2026}{\natexlab{a}})}\BibitemShut {NoStop}%
\bibitem [{\citenamefont {Nolan}\ \emph {et~al.}(2021{\natexlab{a}})\citenamefont {Nolan}, \citenamefont {Smerzi},\ and\ \citenamefont {Pezz{\`e}}}]{nolan2021machine}%
  \BibitemOpen
  \bibfield  {author} {\bibinfo {author} {\bibfnamefont {S.}~\bibnamefont {Nolan}}, \bibinfo {author} {\bibfnamefont {A.}~\bibnamefont {Smerzi}},\ and\ \bibinfo {author} {\bibfnamefont {L.}~\bibnamefont {Pezz{\`e}}},\ }\href {https://doi.org/10.1038/s41534-021-00497-w} {\bibfield  {journal} {\bibinfo  {journal} {npj Quantum Information}\ }\textbf {\bibinfo {volume} {7}},\ \bibinfo {pages} {169} (\bibinfo {year} {2021}{\natexlab{a}})}\BibitemShut {NoStop}%
\bibitem [{\citenamefont {Nolan}\ \emph {et~al.}(2021{\natexlab{b}})\citenamefont {Nolan}, \citenamefont {Pezzè},\ and\ \citenamefont {Smerzi}}]{nolan2021frequentist}%
  \BibitemOpen
  \bibfield  {author} {\bibinfo {author} {\bibfnamefont {S.~P.}\ \bibnamefont {Nolan}}, \bibinfo {author} {\bibfnamefont {L.}~\bibnamefont {Pezzè}},\ and\ \bibinfo {author} {\bibfnamefont {A.}~\bibnamefont {Smerzi}},\ }\href {https://doi.org/10.1116/5.0058163} {\bibfield  {journal} {\bibinfo  {journal} {AVS Quantum Science}\ }\textbf {\bibinfo {volume} {3}},\ \bibinfo {pages} {034401} (\bibinfo {year} {2021}{\natexlab{b}})}\BibitemShut {NoStop}%
\bibitem [{\citenamefont {Huang}\ \emph {et~al.}(2025)\citenamefont {Huang}, \citenamefont {Zhuang}, \citenamefont {Zhou}, \citenamefont {Shen},\ and\ \citenamefont {Lee}}]{huang2025quantum}%
  \BibitemOpen
  \bibfield  {author} {\bibinfo {author} {\bibfnamefont {J.}~\bibnamefont {Huang}}, \bibinfo {author} {\bibfnamefont {M.}~\bibnamefont {Zhuang}}, \bibinfo {author} {\bibfnamefont {J.}~\bibnamefont {Zhou}}, \bibinfo {author} {\bibfnamefont {Y.}~\bibnamefont {Shen}},\ and\ \bibinfo {author} {\bibfnamefont {C.}~\bibnamefont {Lee}},\ }\href {https://doi.org/https://doi.org/10.1002/qute.202300329} {\bibfield  {journal} {\bibinfo  {journal} {Advanced Quantum Technologies}\ }\textbf {\bibinfo {volume} {8}},\ \bibinfo {pages} {2300329} (\bibinfo {year} {2025})}\BibitemShut {NoStop}%
\bibitem [{\citenamefont {Xiao}\ \emph {et~al.}(2022)\citenamefont {Xiao}, \citenamefont {Fan},\ and\ \citenamefont {Zeng}}]{xiao2022parameter}%
  \BibitemOpen
  \bibfield  {author} {\bibinfo {author} {\bibfnamefont {T.}~\bibnamefont {Xiao}}, \bibinfo {author} {\bibfnamefont {J.}~\bibnamefont {Fan}},\ and\ \bibinfo {author} {\bibfnamefont {G.}~\bibnamefont {Zeng}},\ }\href {https://doi.org/10.1038/s41534-021-00513-z} {\bibfield  {journal} {\bibinfo  {journal} {npj Quantum Information}\ }\textbf {\bibinfo {volume} {8}},\ \bibinfo {pages} {2} (\bibinfo {year} {2022})}\BibitemShut {NoStop}%
\bibitem [{\citenamefont {Schuff}\ \emph {et~al.}(2020)\citenamefont {Schuff}, \citenamefont {Fiderer},\ and\ \citenamefont {Braun}}]{schuff2020improving}%
  \BibitemOpen
  \bibfield  {author} {\bibinfo {author} {\bibfnamefont {J.}~\bibnamefont {Schuff}}, \bibinfo {author} {\bibfnamefont {L.~J.}\ \bibnamefont {Fiderer}},\ and\ \bibinfo {author} {\bibfnamefont {D.}~\bibnamefont {Braun}},\ }\href {https://doi.org/10.1088/1367-2630/ab6f1f} {\bibfield  {journal} {\bibinfo  {journal} {New Journal of Physics}\ }\textbf {\bibinfo {volume} {22}},\ \bibinfo {pages} {035001} (\bibinfo {year} {2020})}\BibitemShut {NoStop}%
\bibitem [{\citenamefont {Costa}\ \emph {et~al.}(2021)\citenamefont {Costa}, \citenamefont {Omar}, \citenamefont {Sultanov},\ and\ \citenamefont {Paraoanu}}]{costa2021benchmarking}%
  \BibitemOpen
  \bibfield  {author} {\bibinfo {author} {\bibfnamefont {N.~F.}\ \bibnamefont {Costa}}, \bibinfo {author} {\bibfnamefont {Y.}~\bibnamefont {Omar}}, \bibinfo {author} {\bibfnamefont {A.}~\bibnamefont {Sultanov}},\ and\ \bibinfo {author} {\bibfnamefont {G.~S.}\ \bibnamefont {Paraoanu}},\ }\href {https://doi.org/10.1140/epjqt/s40507-021-00105-y} {\bibfield  {journal} {\bibinfo  {journal} {EPJ Quantum Technology}\ }\textbf {\bibinfo {volume} {8}},\ \bibinfo {pages} {16} (\bibinfo {year} {2021})}\BibitemShut {NoStop}%
\bibitem [{\citenamefont {Mujal}\ \emph {et~al.}(2021)\citenamefont {Mujal}, \citenamefont {Martínez-Peña}, \citenamefont {Nokkala}, \citenamefont {García-Beni}, \citenamefont {Giorgi}, \citenamefont {Soriano},\ and\ \citenamefont {Zambrini}}]{mujal2021opportunities}%
  \BibitemOpen
  \bibfield  {author} {\bibinfo {author} {\bibfnamefont {P.}~\bibnamefont {Mujal}}, \bibinfo {author} {\bibfnamefont {R.}~\bibnamefont {Martínez-Peña}}, \bibinfo {author} {\bibfnamefont {J.}~\bibnamefont {Nokkala}}, \bibinfo {author} {\bibfnamefont {J.}~\bibnamefont {García-Beni}}, \bibinfo {author} {\bibfnamefont {G.~L.}\ \bibnamefont {Giorgi}}, \bibinfo {author} {\bibfnamefont {M.~C.}\ \bibnamefont {Soriano}},\ and\ \bibinfo {author} {\bibfnamefont {R.}~\bibnamefont {Zambrini}},\ }\href {https://doi.org/https://doi.org/10.1002/qute.202100027} {\bibfield  {journal} {\bibinfo  {journal} {Advanced Quantum Technologies}\ }\textbf {\bibinfo {volume} {4}},\ \bibinfo {pages} {2100027} (\bibinfo {year} {2021})}\BibitemShut {NoStop}%
\bibitem [{\citenamefont {Innocenti}\ \emph {et~al.}(2023)\citenamefont {Innocenti}, \citenamefont {Lorenzo}, \citenamefont {Palmisano}, \citenamefont {Ferraro}, \citenamefont {Paternostro},\ and\ \citenamefont {Palma}}]{innocenti2023potential}%
  \BibitemOpen
  \bibfield  {author} {\bibinfo {author} {\bibfnamefont {L.}~\bibnamefont {Innocenti}}, \bibinfo {author} {\bibfnamefont {S.}~\bibnamefont {Lorenzo}}, \bibinfo {author} {\bibfnamefont {I.}~\bibnamefont {Palmisano}}, \bibinfo {author} {\bibfnamefont {A.}~\bibnamefont {Ferraro}}, \bibinfo {author} {\bibfnamefont {M.}~\bibnamefont {Paternostro}},\ and\ \bibinfo {author} {\bibfnamefont {G.~M.}\ \bibnamefont {Palma}},\ }\href {https://doi.org/10.1038/s42005-023-01233-w} {\bibfield  {journal} {\bibinfo  {journal} {Communications Physics}\ }\textbf {\bibinfo {volume} {6}},\ \bibinfo {pages} {118} (\bibinfo {year} {2023})}\BibitemShut {NoStop}%
\bibitem [{\citenamefont {Zia}\ \emph {et~al.}(2025)\citenamefont {Zia}, \citenamefont {Innocenti}, \citenamefont {Minati}, \citenamefont {Lorenzo}, \citenamefont {Suprano}, \citenamefont {Bartolo}, \citenamefont {Spagnolo}, \citenamefont {Giordani}, \citenamefont {Cimini}, \citenamefont {Palma}, \citenamefont {Ferraro}, \citenamefont {Sciarrino},\ and\ \citenamefont {Paternostro}}]{zia2025quantum}%
  \BibitemOpen
  \bibfield  {author} {\bibinfo {author} {\bibfnamefont {D.}~\bibnamefont {Zia}}, \bibinfo {author} {\bibfnamefont {L.}~\bibnamefont {Innocenti}}, \bibinfo {author} {\bibfnamefont {G.}~\bibnamefont {Minati}}, \bibinfo {author} {\bibfnamefont {S.}~\bibnamefont {Lorenzo}}, \bibinfo {author} {\bibfnamefont {A.}~\bibnamefont {Suprano}}, \bibinfo {author} {\bibfnamefont {R.~D.}\ \bibnamefont {Bartolo}}, \bibinfo {author} {\bibfnamefont {N.}~\bibnamefont {Spagnolo}}, \bibinfo {author} {\bibfnamefont {T.}~\bibnamefont {Giordani}}, \bibinfo {author} {\bibfnamefont {V.}~\bibnamefont {Cimini}}, \bibinfo {author} {\bibfnamefont {G.~M.}\ \bibnamefont {Palma}}, \bibinfo {author} {\bibfnamefont {A.}~\bibnamefont {Ferraro}}, \bibinfo {author} {\bibfnamefont {F.}~\bibnamefont {Sciarrino}},\ and\ \bibinfo {author} {\bibfnamefont {M.}~\bibnamefont {Paternostro}},\ }\href {https://doi.org/10.1126/sciadv.ady7987} {\bibfield  {journal} {\bibinfo  {journal} {Science Advances}\ }\textbf {\bibinfo {volume} {11}},\ \bibinfo {pages}
  {eady7987} (\bibinfo {year} {2025})}\BibitemShut {NoStop}%
\bibitem [{\citenamefont {Suprano}\ \emph {et~al.}(2024)\citenamefont {Suprano}, \citenamefont {Zia}, \citenamefont {Innocenti}, \citenamefont {Lorenzo}, \citenamefont {Cimini}, \citenamefont {Giordani}, \citenamefont {Palmisano}, \citenamefont {Polino}, \citenamefont {Spagnolo}, \citenamefont {Sciarrino}, \citenamefont {Palma}, \citenamefont {Ferraro},\ and\ \citenamefont {Paternostro}}]{garcia2023scalable}%
  \BibitemOpen
  \bibfield  {author} {\bibinfo {author} {\bibfnamefont {A.}~\bibnamefont {Suprano}}, \bibinfo {author} {\bibfnamefont {D.}~\bibnamefont {Zia}}, \bibinfo {author} {\bibfnamefont {L.}~\bibnamefont {Innocenti}}, \bibinfo {author} {\bibfnamefont {S.}~\bibnamefont {Lorenzo}}, \bibinfo {author} {\bibfnamefont {V.}~\bibnamefont {Cimini}}, \bibinfo {author} {\bibfnamefont {T.}~\bibnamefont {Giordani}}, \bibinfo {author} {\bibfnamefont {I.}~\bibnamefont {Palmisano}}, \bibinfo {author} {\bibfnamefont {E.}~\bibnamefont {Polino}}, \bibinfo {author} {\bibfnamefont {N.}~\bibnamefont {Spagnolo}}, \bibinfo {author} {\bibfnamefont {F.}~\bibnamefont {Sciarrino}}, \bibinfo {author} {\bibfnamefont {G.~M.}\ \bibnamefont {Palma}}, \bibinfo {author} {\bibfnamefont {A.}~\bibnamefont {Ferraro}},\ and\ \bibinfo {author} {\bibfnamefont {M.}~\bibnamefont {Paternostro}},\ }\href {https://doi.org/10.1103/PhysRevLett.132.160802} {\bibfield  {journal} {\bibinfo  {journal} {Phys. Rev. Lett.}\ }\textbf {\bibinfo {volume} {132}},\ \bibinfo
  {pages} {160802} (\bibinfo {year} {2024})}\BibitemShut {NoStop}%
\bibitem [{\citenamefont {Mujal}\ \emph {et~al.}(2023)\citenamefont {Mujal}, \citenamefont {Mart{\'\i}nez-Pe{\~n}a}, \citenamefont {Giorgi}, \citenamefont {Soriano},\ and\ \citenamefont {Zambrini}}]{mujal2023time}%
  \BibitemOpen
  \bibfield  {author} {\bibinfo {author} {\bibfnamefont {P.}~\bibnamefont {Mujal}}, \bibinfo {author} {\bibfnamefont {R.}~\bibnamefont {Mart{\'\i}nez-Pe{\~n}a}}, \bibinfo {author} {\bibfnamefont {G.~L.}\ \bibnamefont {Giorgi}}, \bibinfo {author} {\bibfnamefont {M.~C.}\ \bibnamefont {Soriano}},\ and\ \bibinfo {author} {\bibfnamefont {R.}~\bibnamefont {Zambrini}},\ }\href {https://doi.org/10.1038/s41534-023-00682-z} {\bibfield  {journal} {\bibinfo  {journal} {npj Quantum Information}\ }\textbf {\bibinfo {volume} {9}},\ \bibinfo {pages} {16} (\bibinfo {year} {2023})}\BibitemShut {NoStop}%
\bibitem [{\citenamefont {Das}\ \emph {et~al.}(2026)\citenamefont {Das}, \citenamefont {Giorgi},\ and\ \citenamefont {Zambrini}}]{das2026quantum}%
  \BibitemOpen
  \bibfield  {author} {\bibinfo {author} {\bibfnamefont {S.}~\bibnamefont {Das}}, \bibinfo {author} {\bibfnamefont {G.~L.}\ \bibnamefont {Giorgi}},\ and\ \bibinfo {author} {\bibfnamefont {R.}~\bibnamefont {Zambrini}},\ }\href {https://doi.org/10.1103/ffd3-ytbt} {\bibfield  {journal} {\bibinfo  {journal} {Phys. Rev. Res.}\ }\textbf {\bibinfo {volume} {8}},\ \bibinfo {pages} {023148} (\bibinfo {year} {2026})}\BibitemShut {NoStop}%
\bibitem [{\citenamefont {Li}\ \emph {et~al.}(2026{\natexlab{b}})\citenamefont {Li}, \citenamefont {Mukhopadhyay}, \citenamefont {Bayat},\ and\ \citenamefont {Habibnia}}]{li2026quantum}%
  \BibitemOpen
  \bibfield  {author} {\bibinfo {author} {\bibfnamefont {Q.}~\bibnamefont {Li}}, \bibinfo {author} {\bibfnamefont {C.}~\bibnamefont {Mukhopadhyay}}, \bibinfo {author} {\bibfnamefont {A.}~\bibnamefont {Bayat}},\ and\ \bibinfo {author} {\bibfnamefont {A.}~\bibnamefont {Habibnia}},\ }\href {https://doi.org/10.1103/rbj7-4wnq} {\bibfield  {journal} {\bibinfo  {journal} {Phys. Rev. Res.}\ }\textbf {\bibinfo {volume} {8}},\ \bibinfo {pages} {023028} (\bibinfo {year} {2026}{\natexlab{b}})}\BibitemShut {NoStop}%
\bibitem [{\citenamefont {Kobayashi}\ \emph {et~al.}(2024{\natexlab{a}})\citenamefont {Kobayashi}, \citenamefont {Fujii},\ and\ \citenamefont {Yamamoto}}]{kobayashi2024feedback}%
  \BibitemOpen
  \bibfield  {author} {\bibinfo {author} {\bibfnamefont {K.}~\bibnamefont {Kobayashi}}, \bibinfo {author} {\bibfnamefont {K.}~\bibnamefont {Fujii}},\ and\ \bibinfo {author} {\bibfnamefont {N.}~\bibnamefont {Yamamoto}},\ }\href {https://doi.org/10.1103/PRXQuantum.5.040325} {\bibfield  {journal} {\bibinfo  {journal} {PRX Quantum}\ }\textbf {\bibinfo {volume} {5}},\ \bibinfo {pages} {040325} (\bibinfo {year} {2024}{\natexlab{a}})}\BibitemShut {NoStop}%
\bibitem [{\citenamefont {Vetrano}\ \emph {et~al.}(2025)\citenamefont {Vetrano}, \citenamefont {Lo~Monaco}, \citenamefont {Innocenti}, \citenamefont {Lorenzo},\ and\ \citenamefont {Palma}}]{Vetrano_2025}%
  \BibitemOpen
  \bibfield  {author} {\bibinfo {author} {\bibfnamefont {M.}~\bibnamefont {Vetrano}}, \bibinfo {author} {\bibfnamefont {G.}~\bibnamefont {Lo~Monaco}}, \bibinfo {author} {\bibfnamefont {L.}~\bibnamefont {Innocenti}}, \bibinfo {author} {\bibfnamefont {S.}~\bibnamefont {Lorenzo}},\ and\ \bibinfo {author} {\bibfnamefont {G.~M.}\ \bibnamefont {Palma}},\ }\href {https://doi.org/10.1038/s41534-024-00927-5} {\bibfield  {journal} {\bibinfo  {journal} {npj Quantum Information}\ }\textbf {\bibinfo {volume} {11}},\ \bibinfo {pages} {20} (\bibinfo {year} {2025})}\BibitemShut {NoStop}%
\bibitem [{\citenamefont {Liu}\ \emph {et~al.}(2026)\citenamefont {Liu}, \citenamefont {Ren}, \citenamefont {Nill}, \citenamefont {Cabot}, \citenamefont {Xia}, \citenamefont {Tong}, \citenamefont {Wang}, \citenamefont {Yang}, \citenamefont {Xie}, \citenamefont {Jing}, \citenamefont {Zhang}, \citenamefont {Xiao}, \citenamefont {Jia}, \citenamefont {Lesanovsky},\ and\ \citenamefont {Zhang}}]{liu2026time}%
  \BibitemOpen
  \bibfield  {author} {\bibinfo {author} {\bibfnamefont {Z.-K.}\ \bibnamefont {Liu}}, \bibinfo {author} {\bibfnamefont {Q.-M.}\ \bibnamefont {Ren}}, \bibinfo {author} {\bibfnamefont {C.}~\bibnamefont {Nill}}, \bibinfo {author} {\bibfnamefont {A.}~\bibnamefont {Cabot}}, \bibinfo {author} {\bibfnamefont {W.}~\bibnamefont {Xia}}, \bibinfo {author} {\bibfnamefont {Y.-J.}\ \bibnamefont {Tong}}, \bibinfo {author} {\bibfnamefont {H.-Z.}\ \bibnamefont {Wang}}, \bibinfo {author} {\bibfnamefont {W.-W.}\ \bibnamefont {Yang}}, \bibinfo {author} {\bibfnamefont {J.-Y.}\ \bibnamefont {Xie}}, \bibinfo {author} {\bibfnamefont {M.-Y.}\ \bibnamefont {Jing}}, \bibinfo {author} {\bibfnamefont {H.}~\bibnamefont {Zhang}}, \bibinfo {author} {\bibfnamefont {L.-T.}\ \bibnamefont {Xiao}}, \bibinfo {author} {\bibfnamefont {S.-T.}\ \bibnamefont {Jia}}, \bibinfo {author} {\bibfnamefont {I.}~\bibnamefont {Lesanovsky}},\ and\ \bibinfo {author} {\bibfnamefont {L.-J.}\ \bibnamefont {Zhang}},\ }\href
  {https://doi.org/10.1038/s41467-026-77083-1} {\bibfield  {journal} {\bibinfo  {journal} {Nature Communications}\ }\textbf {\bibinfo {volume} {17}},\ \bibinfo {pages} {10118} (\bibinfo {year} {2026})}\BibitemShut {NoStop}%
\bibitem [{\citenamefont {Sannia}\ \emph {et~al.}(2026)\citenamefont {Sannia}, \citenamefont {Menta}, \citenamefont {Sathe}, \citenamefont {Santis}, \citenamefont {Giovannetti}, \citenamefont {García-Pintos}, \citenamefont {Venturelli}, \citenamefont {Giorgi}, \citenamefont {Zambrini},\ and\ \citenamefont {Caravelli}}]{sannia2026temporalinformationprocessing4500qubit}%
  \BibitemOpen
  \bibfield  {author} {\bibinfo {author} {\bibfnamefont {A.}~\bibnamefont {Sannia}}, \bibinfo {author} {\bibfnamefont {R.}~\bibnamefont {Menta}}, \bibinfo {author} {\bibfnamefont {P.}~\bibnamefont {Sathe}}, \bibinfo {author} {\bibfnamefont {D.~D.}\ \bibnamefont {Santis}}, \bibinfo {author} {\bibfnamefont {V.}~\bibnamefont {Giovannetti}}, \bibinfo {author} {\bibfnamefont {L.~P.}\ \bibnamefont {García-Pintos}}, \bibinfo {author} {\bibfnamefont {D.}~\bibnamefont {Venturelli}}, \bibinfo {author} {\bibfnamefont {G.~L.}\ \bibnamefont {Giorgi}}, \bibinfo {author} {\bibfnamefont {R.}~\bibnamefont {Zambrini}},\ and\ \bibinfo {author} {\bibfnamefont {F.}~\bibnamefont {Caravelli}},\ }\href {https://arxiv.org/abs/2609.19308} {\bibinfo {title} {Temporal information processing on a 4,500-qubit quantum annealer}} (\bibinfo {year} {2026}),\ \Eprint {https://arxiv.org/abs/2609.19308} {arXiv:2609.19308 [quant-ph]} \BibitemShut {NoStop}%
\bibitem [{\citenamefont {Xia}\ \emph {et~al.}(2022)\citenamefont {Xia}, \citenamefont {Zou}, \citenamefont {Qiu},\ and\ \citenamefont {Li}}]{xia2022reservoir}%
  \BibitemOpen
  \bibfield  {author} {\bibinfo {author} {\bibfnamefont {W.}~\bibnamefont {Xia}}, \bibinfo {author} {\bibfnamefont {J.}~\bibnamefont {Zou}}, \bibinfo {author} {\bibfnamefont {X.}~\bibnamefont {Qiu}},\ and\ \bibinfo {author} {\bibfnamefont {X.}~\bibnamefont {Li}},\ }\href {https://doi.org/10.1007/s11467-022-1158-1} {\bibfield  {journal} {\bibinfo  {journal} {Frontiers of Physics}\ }\textbf {\bibinfo {volume} {17}},\ \bibinfo {pages} {33506} (\bibinfo {year} {2022})}\BibitemShut {NoStop}%
\bibitem [{\citenamefont {Settino}\ \emph {et~al.}(2025)\citenamefont {Settino}, \citenamefont {Salatino}, \citenamefont {Mariani}, \citenamefont {D'Amore}, \citenamefont {Channab}, \citenamefont {Bozzolo}, \citenamefont {Vallisa}, \citenamefont {Barill{\`a}}, \citenamefont {Policicchio}, \citenamefont {Lo~Gullo}, \citenamefont {Giordano}, \citenamefont {Mastroianni},\ and\ \citenamefont {Plastina}}]{settino2025memory}%
  \BibitemOpen
  \bibfield  {author} {\bibinfo {author} {\bibfnamefont {J.}~\bibnamefont {Settino}}, \bibinfo {author} {\bibfnamefont {L.}~\bibnamefont {Salatino}}, \bibinfo {author} {\bibfnamefont {L.}~\bibnamefont {Mariani}}, \bibinfo {author} {\bibfnamefont {F.}~\bibnamefont {D'Amore}}, \bibinfo {author} {\bibfnamefont {M.}~\bibnamefont {Channab}}, \bibinfo {author} {\bibfnamefont {L.}~\bibnamefont {Bozzolo}}, \bibinfo {author} {\bibfnamefont {S.}~\bibnamefont {Vallisa}}, \bibinfo {author} {\bibfnamefont {P.}~\bibnamefont {Barill{\`a}}}, \bibinfo {author} {\bibfnamefont {A.}~\bibnamefont {Policicchio}}, \bibinfo {author} {\bibfnamefont {N.}~\bibnamefont {Lo~Gullo}}, \bibinfo {author} {\bibfnamefont {A.}~\bibnamefont {Giordano}}, \bibinfo {author} {\bibfnamefont {C.}~\bibnamefont {Mastroianni}},\ and\ \bibinfo {author} {\bibfnamefont {F.}~\bibnamefont {Plastina}},\ }\href {https://doi.org/10.1103/wzwv-7rk2} {\bibfield  {journal} {\bibinfo  {journal} {Phys. Rev. Appl.}\ }\textbf {\bibinfo {volume} {24}},\ \bibinfo {pages}
  {024019} (\bibinfo {year} {2025})}\BibitemShut {NoStop}%
\bibitem [{\citenamefont {Henaff}\ \emph {et~al.}(2024)\citenamefont {Henaff}, \citenamefont {Ansquer}, \citenamefont {Soriano}, \citenamefont {Zambrini}, \citenamefont {Treps},\ and\ \citenamefont {Parigi}}]{henaff2024optical}%
  \BibitemOpen
  \bibfield  {author} {\bibinfo {author} {\bibfnamefont {J.}~\bibnamefont {Henaff}}, \bibinfo {author} {\bibfnamefont {M.}~\bibnamefont {Ansquer}}, \bibinfo {author} {\bibfnamefont {M.~C.}\ \bibnamefont {Soriano}}, \bibinfo {author} {\bibfnamefont {R.}~\bibnamefont {Zambrini}}, \bibinfo {author} {\bibfnamefont {N.}~\bibnamefont {Treps}},\ and\ \bibinfo {author} {\bibfnamefont {V.}~\bibnamefont {Parigi}},\ }\href {https://doi.org/10.1364/OL.518505} {\bibfield  {journal} {\bibinfo  {journal} {Opt. Lett.}\ }\textbf {\bibinfo {volume} {49}},\ \bibinfo {pages} {2097} (\bibinfo {year} {2024})}\BibitemShut {NoStop}%
\bibitem [{\citenamefont {Hou}\ \emph {et~al.}(2026)\citenamefont {Hou}, \citenamefont {Hua}, \citenamefont {Wu}, \citenamefont {Xia}, \citenamefont {Chen}, \citenamefont {Li}, \citenamefont {Li}, \citenamefont {Peng},\ and\ \citenamefont {Du}}]{hou2026high}%
  \BibitemOpen
  \bibfield  {author} {\bibinfo {author} {\bibfnamefont {Y.}~\bibnamefont {Hou}}, \bibinfo {author} {\bibfnamefont {J.}~\bibnamefont {Hua}}, \bibinfo {author} {\bibfnamefont {Z.}~\bibnamefont {Wu}}, \bibinfo {author} {\bibfnamefont {W.}~\bibnamefont {Xia}}, \bibinfo {author} {\bibfnamefont {Y.}~\bibnamefont {Chen}}, \bibinfo {author} {\bibfnamefont {X.}~\bibnamefont {Li}}, \bibinfo {author} {\bibfnamefont {Z.}~\bibnamefont {Li}}, \bibinfo {author} {\bibfnamefont {X.}~\bibnamefont {Peng}},\ and\ \bibinfo {author} {\bibfnamefont {J.}~\bibnamefont {Du}},\ }\href {https://doi.org/10.1103/r8ww-qw7j} {\bibfield  {journal} {\bibinfo  {journal} {Phys. Rev. Lett.}\ }\textbf {\bibinfo {volume} {136}},\ \bibinfo {pages} {120602} (\bibinfo {year} {2026})}\BibitemShut {NoStop}%
\bibitem [{\citenamefont {Llodrà}\ \emph {et~al.}(2025)\citenamefont {Llodrà}, \citenamefont {Mujal}, \citenamefont {Zambrini},\ and\ \citenamefont {Giorgi}}]{llodra2025quantum}%
  \BibitemOpen
  \bibfield  {author} {\bibinfo {author} {\bibfnamefont {G.}~\bibnamefont {Llodrà}}, \bibinfo {author} {\bibfnamefont {P.}~\bibnamefont {Mujal}}, \bibinfo {author} {\bibfnamefont {R.}~\bibnamefont {Zambrini}},\ and\ \bibinfo {author} {\bibfnamefont {G.~L.}\ \bibnamefont {Giorgi}},\ }\href {https://doi.org/https://doi.org/10.1016/j.chaos.2025.116289} {\bibfield  {journal} {\bibinfo  {journal} {Chaos, Solitons Fractals}\ }\textbf {\bibinfo {volume} {195}},\ \bibinfo {pages} {116289} (\bibinfo {year} {2025})}\BibitemShut {NoStop}%
\bibitem [{\citenamefont {Sharifian}\ \emph {et~al.}(2026)\citenamefont {Sharifian}, \citenamefont {Li}, \citenamefont {Mukhopadhyay}, \citenamefont {Bayat},\ and\ \citenamefont {Minati}}]{sharifian2026quantum}%
  \BibitemOpen
  \bibfield  {author} {\bibinfo {author} {\bibfnamefont {M.}~\bibnamefont {Sharifian}}, \bibinfo {author} {\bibfnamefont {Q.}~\bibnamefont {Li}}, \bibinfo {author} {\bibfnamefont {C.}~\bibnamefont {Mukhopadhyay}}, \bibinfo {author} {\bibfnamefont {A.}~\bibnamefont {Bayat}},\ and\ \bibinfo {author} {\bibfnamefont {L.}~\bibnamefont {Minati}},\ }\href {https://arxiv.org/abs/2610.13388} {\bibinfo {title} {Quantum extreme learning for forecasting experimental chaotic electronic circuits}} (\bibinfo {year} {2026}),\ \Eprint {https://arxiv.org/abs/2610.13388} {arXiv:2610.13388 [quant-ph]} \BibitemShut {NoStop}%
\bibitem [{\citenamefont {Zhang}\ \emph {et~al.}(2026)\citenamefont {Zhang}, \citenamefont {Li}, \citenamefont {Guo}, \citenamefont {Yu}, \citenamefont {Jin},\ and\ \citenamefont {Yin}}]{zhang2026robust}%
  \BibitemOpen
  \bibfield  {author} {\bibinfo {author} {\bibfnamefont {D.}~\bibnamefont {Zhang}}, \bibinfo {author} {\bibfnamefont {X.}~\bibnamefont {Li}}, \bibinfo {author} {\bibfnamefont {Y.}~\bibnamefont {Guo}}, \bibinfo {author} {\bibfnamefont {H.}~\bibnamefont {Yu}}, \bibinfo {author} {\bibfnamefont {Y.}~\bibnamefont {Jin}},\ and\ \bibinfo {author} {\bibfnamefont {Z.-Q.}\ \bibnamefont {Yin}},\ }\href {https://doi.org/10.1103/w2w9-c944} {\bibfield  {journal} {\bibinfo  {journal} {Phys. Rev. Appl.}\ }\textbf {\bibinfo {volume} {26}},\ \bibinfo {pages} {014056} (\bibinfo {year} {2026})}\BibitemShut {NoStop}%
\bibitem [{\citenamefont {Cerezo}\ \emph {et~al.}(2021)\citenamefont {Cerezo}, \citenamefont {Arrasmith}, \citenamefont {Babbush}, \citenamefont {Benjamin}, \citenamefont {Endo}, \citenamefont {Fujii}, \citenamefont {McClean}, \citenamefont {Mitarai}, \citenamefont {Yuan}, \citenamefont {Cincio} \emph {et~al.}}]{cerezo2021variational}%
  \BibitemOpen
  \bibfield  {author} {\bibinfo {author} {\bibfnamefont {M.}~\bibnamefont {Cerezo}}, \bibinfo {author} {\bibfnamefont {A.}~\bibnamefont {Arrasmith}}, \bibinfo {author} {\bibfnamefont {R.}~\bibnamefont {Babbush}}, \bibinfo {author} {\bibfnamefont {S.~C.}\ \bibnamefont {Benjamin}}, \bibinfo {author} {\bibfnamefont {S.}~\bibnamefont {Endo}}, \bibinfo {author} {\bibfnamefont {K.}~\bibnamefont {Fujii}}, \bibinfo {author} {\bibfnamefont {J.~R.}\ \bibnamefont {McClean}}, \bibinfo {author} {\bibfnamefont {K.}~\bibnamefont {Mitarai}}, \bibinfo {author} {\bibfnamefont {X.}~\bibnamefont {Yuan}}, \bibinfo {author} {\bibfnamefont {L.}~\bibnamefont {Cincio}}, \emph {et~al.},\ }\href {https://doi.org/10.1038/s42254-021-00348-9} {\bibfield  {journal} {\bibinfo  {journal} {Nature Reviews Physics}\ }\textbf {\bibinfo {volume} {3}},\ \bibinfo {pages} {625} (\bibinfo {year} {2021})}\BibitemShut {NoStop}%
\bibitem [{Note1()}]{Note1}%
  \BibitemOpen
  \bibinfo {note} {Note that, in QEL, quantum reservoir refers to the quantum system used to encode and process the input data, rather than to an environmental bath.}\BibitemShut {Stop}%
\bibitem [{\citenamefont {Fujii}\ and\ \citenamefont {Nakajima}(2017)}]{fujii2017}%
  \BibitemOpen
  \bibfield  {author} {\bibinfo {author} {\bibfnamefont {K.}~\bibnamefont {Fujii}}\ and\ \bibinfo {author} {\bibfnamefont {K.}~\bibnamefont {Nakajima}},\ }\href {https://doi.org/10.1103/PhysRevApplied.8.024030} {\bibfield  {journal} {\bibinfo  {journal} {Phys. Rev. Appl.}\ }\textbf {\bibinfo {volume} {8}},\ \bibinfo {pages} {024030} (\bibinfo {year} {2017})}\BibitemShut {NoStop}%
\bibitem [{\citenamefont {Ghosh}\ \emph {et~al.}(2021)\citenamefont {Ghosh}, \citenamefont {Nakajima}, \citenamefont {Krisnanda}, \citenamefont {Fujii},\ and\ \citenamefont {Liew}}]{ghosh2021}%
  \BibitemOpen
  \bibfield  {author} {\bibinfo {author} {\bibfnamefont {S.}~\bibnamefont {Ghosh}}, \bibinfo {author} {\bibfnamefont {K.}~\bibnamefont {Nakajima}}, \bibinfo {author} {\bibfnamefont {T.}~\bibnamefont {Krisnanda}}, \bibinfo {author} {\bibfnamefont {K.}~\bibnamefont {Fujii}},\ and\ \bibinfo {author} {\bibfnamefont {T.~C.~H.}\ \bibnamefont {Liew}},\ }\href {https://doi.org/https://doi.org/10.1002/qute.202100053} {\bibfield  {journal} {\bibinfo  {journal} {Advanced Quantum Technologies}\ }\textbf {\bibinfo {volume} {4}},\ \bibinfo {pages} {2100053} (\bibinfo {year} {2021})}\BibitemShut {NoStop}%
\bibitem [{\citenamefont {Kobayashi}\ \emph {et~al.}(2024{\natexlab{b}})\citenamefont {Kobayashi}, \citenamefont {Fujii},\ and\ \citenamefont {Yamamoto}}]{kaito2024}%
  \BibitemOpen
  \bibfield  {author} {\bibinfo {author} {\bibfnamefont {K.}~\bibnamefont {Kobayashi}}, \bibinfo {author} {\bibfnamefont {K.}~\bibnamefont {Fujii}},\ and\ \bibinfo {author} {\bibfnamefont {N.}~\bibnamefont {Yamamoto}},\ }\href {https://doi.org/10.1103/PRXQuantum.5.040325} {\bibfield  {journal} {\bibinfo  {journal} {PRX Quantum}\ }\textbf {\bibinfo {volume} {5}},\ \bibinfo {pages} {040325} (\bibinfo {year} {2024}{\natexlab{b}})}\BibitemShut {NoStop}%
\bibitem [{\citenamefont {Barouch}\ \emph {et~al.}(1970)\citenamefont {Barouch}, \citenamefont {McCoy},\ and\ \citenamefont {Dresden}}]{barouch1970}%
  \BibitemOpen
  \bibfield  {author} {\bibinfo {author} {\bibfnamefont {E.}~\bibnamefont {Barouch}}, \bibinfo {author} {\bibfnamefont {B.~M.}\ \bibnamefont {McCoy}},\ and\ \bibinfo {author} {\bibfnamefont {M.}~\bibnamefont {Dresden}},\ }\href {https://doi.org/10.1103/PhysRevA.2.1075} {\bibfield  {journal} {\bibinfo  {journal} {Phys. Rev. A}\ }\textbf {\bibinfo {volume} {2}},\ \bibinfo {pages} {1075} (\bibinfo {year} {1970})}\BibitemShut {NoStop}%
\bibitem [{\citenamefont {Mbeng}\ \emph {et~al.}(2024)\citenamefont {Mbeng}, \citenamefont {Russomanno},\ and\ \citenamefont {Santoro}}]{santoro2024}%
  \BibitemOpen
  \bibfield  {author} {\bibinfo {author} {\bibfnamefont {G.~B.}\ \bibnamefont {Mbeng}}, \bibinfo {author} {\bibfnamefont {A.}~\bibnamefont {Russomanno}},\ and\ \bibinfo {author} {\bibfnamefont {G.~E.}\ \bibnamefont {Santoro}},\ }\href {https://doi.org/10.21468/SciPostPhysLectNotes.82} {\bibfield  {journal} {\bibinfo  {journal} {SciPost Phys. Lect. Notes}\ ,\ \bibinfo {pages} {82}} (\bibinfo {year} {2024})}\BibitemShut {NoStop}%
\bibitem [{\citenamefont {PARIS}(2009)}]{paris2009quantum}%
  \BibitemOpen
  \bibfield  {author} {\bibinfo {author} {\bibfnamefont {M.~G.~A.}\ \bibnamefont {PARIS}},\ }\href {https://doi.org/10.1142/S0219749909004839} {\bibfield  {journal} {\bibinfo  {journal} {International Journal of Quantum Information}\ }\textbf {\bibinfo {volume} {07}},\ \bibinfo {pages} {125} (\bibinfo {year} {2009})}\BibitemShut {NoStop}%
\bibitem [{\citenamefont {Liu}\ \emph {et~al.}(2019)\citenamefont {Liu}, \citenamefont {Yuan}, \citenamefont {Lu},\ and\ \citenamefont {Wang}}]{liu2020quantum}%
  \BibitemOpen
  \bibfield  {author} {\bibinfo {author} {\bibfnamefont {J.}~\bibnamefont {Liu}}, \bibinfo {author} {\bibfnamefont {H.}~\bibnamefont {Yuan}}, \bibinfo {author} {\bibfnamefont {X.-M.}\ \bibnamefont {Lu}},\ and\ \bibinfo {author} {\bibfnamefont {X.}~\bibnamefont {Wang}},\ }\href {https://doi.org/10.1088/1751-8121/ab5d4d} {\bibfield  {journal} {\bibinfo  {journal} {Journal of Physics A: Mathematical and Theoretical}\ }\textbf {\bibinfo {volume} {53}},\ \bibinfo {pages} {023001} (\bibinfo {year} {2019})}\BibitemShut {NoStop}%
\bibitem [{\citenamefont {Meyer}(2021)}]{meyer2021fisher}%
  \BibitemOpen
  \bibfield  {author} {\bibinfo {author} {\bibfnamefont {J.~J.}\ \bibnamefont {Meyer}},\ }\href {https://doi.org/10.22331/q-2021-09-09-539} {\bibfield  {journal} {\bibinfo  {journal} {{Quantum}}\ }\textbf {\bibinfo {volume} {5}},\ \bibinfo {pages} {539} (\bibinfo {year} {2021})}\BibitemShut {NoStop}%
\bibitem [{\citenamefont {Boixo}\ \emph {et~al.}(2007)\citenamefont {Boixo}, \citenamefont {Flammia}, \citenamefont {Caves},\ and\ \citenamefont {Geremia}}]{bixo2007generalized}%
  \BibitemOpen
  \bibfield  {author} {\bibinfo {author} {\bibfnamefont {S.}~\bibnamefont {Boixo}}, \bibinfo {author} {\bibfnamefont {S.~T.}\ \bibnamefont {Flammia}}, \bibinfo {author} {\bibfnamefont {C.~M.}\ \bibnamefont {Caves}},\ and\ \bibinfo {author} {\bibfnamefont {J.}~\bibnamefont {Geremia}},\ }\href {https://doi.org/10.1103/PhysRevLett.98.090401} {\bibfield  {journal} {\bibinfo  {journal} {Phys. Rev. Lett.}\ }\textbf {\bibinfo {volume} {98}},\ \bibinfo {pages} {090401} (\bibinfo {year} {2007})}\BibitemShut {NoStop}%
\bibitem [{\citenamefont {O'Connor}\ \emph {et~al.}(2026)\citenamefont {O'Connor}, \citenamefont {He}, \citenamefont {Paris},\ and\ \citenamefont {Genoni}}]{o2026geometric}%
  \BibitemOpen
  \bibfield  {author} {\bibinfo {author} {\bibfnamefont {E.}~\bibnamefont {O'Connor}}, \bibinfo {author} {\bibfnamefont {J.}~\bibnamefont {He}}, \bibinfo {author} {\bibfnamefont {M.~G.~A.}\ \bibnamefont {Paris}},\ and\ \bibinfo {author} {\bibfnamefont {M.~G.}\ \bibnamefont {Genoni}},\ }\href {https://arxiv.org/abs/2607.06410} {\bibinfo {title} {Geometric obstructions to quadratic time scaling in multiparameter quantum estimation}} (\bibinfo {year} {2026}),\ \Eprint {https://arxiv.org/abs/2607.06410} {arXiv:2607.06410 [quant-ph]} \BibitemShut {NoStop}%
\bibitem [{\citenamefont {Zheng}\ \emph {et~al.}(2026)\citenamefont {Zheng}, \citenamefont {Yang},\ and\ \citenamefont {Bayat}}]{zheng2026stabilizing}%
  \BibitemOpen
  \bibfield  {author} {\bibinfo {author} {\bibfnamefont {W.}~\bibnamefont {Zheng}}, \bibinfo {author} {\bibfnamefont {Y.}~\bibnamefont {Yang}},\ and\ \bibinfo {author} {\bibfnamefont {A.}~\bibnamefont {Bayat}},\ }\href {https://arxiv.org/abs/2609.24805} {\bibinfo {title} {Stabilizing temporal quantum-enhanced sensitivity via sub-optimal measurements}} (\bibinfo {year} {2026}),\ \Eprint {https://arxiv.org/abs/2609.24805} {arXiv:2609.24805 [quant-ph]} \BibitemShut {NoStop}%
\bibitem [{\citenamefont {Thanasilp}\ \emph {et~al.}(2024)\citenamefont {Thanasilp}, \citenamefont {Wang}, \citenamefont {Cerezo},\ and\ \citenamefont {Holmes}}]{thanasilp2024exponential}%
  \BibitemOpen
  \bibfield  {author} {\bibinfo {author} {\bibfnamefont {S.}~\bibnamefont {Thanasilp}}, \bibinfo {author} {\bibfnamefont {S.}~\bibnamefont {Wang}}, \bibinfo {author} {\bibfnamefont {M.}~\bibnamefont {Cerezo}},\ and\ \bibinfo {author} {\bibfnamefont {Z.}~\bibnamefont {Holmes}},\ }\href {https://doi.org/10.1038/s41467-024-49287-w} {\bibfield  {journal} {\bibinfo  {journal} {Nature communications}\ }\textbf {\bibinfo {volume} {15}},\ \bibinfo {pages} {5200} (\bibinfo {year} {2024})}\BibitemShut {NoStop}%
\bibitem [{\citenamefont {Sannia}\ \emph {et~al.}(2025)\citenamefont {Sannia}, \citenamefont {Giorgi},\ and\ \citenamefont {Zambrini}}]{sannia2025exponential}%
  \BibitemOpen
  \bibfield  {author} {\bibinfo {author} {\bibfnamefont {A.}~\bibnamefont {Sannia}}, \bibinfo {author} {\bibfnamefont {G.~L.}\ \bibnamefont {Giorgi}},\ and\ \bibinfo {author} {\bibfnamefont {R.}~\bibnamefont {Zambrini}},\ }\href {https://arxiv.org/abs/2505.10062} {\bibinfo {title} {Exponential concentration and symmetries in quantum reservoir computing}} (\bibinfo {year} {2025}),\ \Eprint {https://arxiv.org/abs/2505.10062} {arXiv:2505.10062 [quant-ph]} \BibitemShut {NoStop}%
\bibitem [{\citenamefont {Xiong}\ \emph {et~al.}(2025)\citenamefont {Xiong}, \citenamefont {Facelli}, \citenamefont {Sahebi}, \citenamefont {Agnel}, \citenamefont {Chotibut}, \citenamefont {Thanasilp},\ and\ \citenamefont {Holmes}}]{xiong2025fundamental}%
  \BibitemOpen
  \bibfield  {author} {\bibinfo {author} {\bibfnamefont {W.}~\bibnamefont {Xiong}}, \bibinfo {author} {\bibfnamefont {G.}~\bibnamefont {Facelli}}, \bibinfo {author} {\bibfnamefont {M.}~\bibnamefont {Sahebi}}, \bibinfo {author} {\bibfnamefont {O.}~\bibnamefont {Agnel}}, \bibinfo {author} {\bibfnamefont {T.}~\bibnamefont {Chotibut}}, \bibinfo {author} {\bibfnamefont {S.}~\bibnamefont {Thanasilp}},\ and\ \bibinfo {author} {\bibfnamefont {Z.}~\bibnamefont {Holmes}},\ }\href {https://doi.org/10.1007/s42484-025-00239-7} {\bibfield  {journal} {\bibinfo  {journal} {Quantum Machine Intelligence}\ }\textbf {\bibinfo {volume} {7}},\ \bibinfo {pages} {20} (\bibinfo {year} {2025})}\BibitemShut {NoStop}%
\bibitem [{\citenamefont {Demkowicz-Dobrzański}\ \emph {et~al.}(2012)\citenamefont {Demkowicz-Dobrzański}, \citenamefont {Kołodyński},\ and\ \citenamefont {Guţă}}]{demkowicz2012elusive}%
  \BibitemOpen
  \bibfield  {author} {\bibinfo {author} {\bibfnamefont {R.}~\bibnamefont {Demkowicz-Dobrzański}}, \bibinfo {author} {\bibfnamefont {J.}~\bibnamefont {Kołodyński}},\ and\ \bibinfo {author} {\bibfnamefont {M.}~\bibnamefont {Guţă}},\ }\href {https://doi.org/10.1038/ncomms2067} {\bibfield  {journal} {\bibinfo  {journal} {Nature Communications}\ }\textbf {\bibinfo {volume} {3}},\ \bibinfo {pages} {1063} (\bibinfo {year} {2012})}\BibitemShut {NoStop}%
\bibitem [{\citenamefont {Kołodyński}\ and\ \citenamefont {Demkowicz-Dobrzański}(2013)}]{Kołodyński_2013}%
  \BibitemOpen
  \bibfield  {author} {\bibinfo {author} {\bibfnamefont {J.}~\bibnamefont {Kołodyński}}\ and\ \bibinfo {author} {\bibfnamefont {R.}~\bibnamefont {Demkowicz-Dobrzański}},\ }\href {https://doi.org/10.1088/1367-2630/15/7/073043} {\bibfield  {journal} {\bibinfo  {journal} {New Journal of Physics}\ }\textbf {\bibinfo {volume} {15}},\ \bibinfo {pages} {073043} (\bibinfo {year} {2013})}\BibitemShut {NoStop}%
\bibitem [{\citenamefont {Alipour}\ \emph {et~al.}(2014)\citenamefont {Alipour}, \citenamefont {Mehboudi},\ and\ \citenamefont {Rezakhani}}]{alipour2014}%
  \BibitemOpen
  \bibfield  {author} {\bibinfo {author} {\bibfnamefont {S.}~\bibnamefont {Alipour}}, \bibinfo {author} {\bibfnamefont {M.}~\bibnamefont {Mehboudi}},\ and\ \bibinfo {author} {\bibfnamefont {A.~T.}\ \bibnamefont {Rezakhani}},\ }\href {https://doi.org/10.1103/PhysRevLett.112.120405} {\bibfield  {journal} {\bibinfo  {journal} {Phys. Rev. Lett.}\ }\textbf {\bibinfo {volume} {112}},\ \bibinfo {pages} {120405} (\bibinfo {year} {2014})}\BibitemShut {NoStop}%
\bibitem [{\citenamefont {Demkowicz-Dobrza{\ifmmode \acute{n}\else \'{n}\fi{}}ski}\ and\ \citenamefont {Maccone}(2014)}]{rafal2014}%
  \BibitemOpen
  \bibfield  {author} {\bibinfo {author} {\bibfnamefont {R.}~\bibnamefont {Demkowicz-Dobrza{\ifmmode \acute{n}\else \'{n}\fi{}}ski}}\ and\ \bibinfo {author} {\bibfnamefont {L.}~\bibnamefont {Maccone}},\ }\href {https://doi.org/10.1103/PhysRevLett.113.250801} {\bibfield  {journal} {\bibinfo  {journal} {Phys. Rev. Lett.}\ }\textbf {\bibinfo {volume} {113}},\ \bibinfo {pages} {250801} (\bibinfo {year} {2014})}\BibitemShut {NoStop}%
\bibitem [{\citenamefont {Ghosh}\ \emph {et~al.}(2026{\natexlab{b}})\citenamefont {Ghosh}, \citenamefont {Kobus}, \citenamefont {Kurdzia{\l{}}ek},\ and\ \citenamefont {Demkowicz-Dobrza{\'{n}}ski}}]{Ghosh2026optimalphase}%
  \BibitemOpen
  \bibfield  {author} {\bibinfo {author} {\bibfnamefont {S.}~\bibnamefont {Ghosh}}, \bibinfo {author} {\bibfnamefont {A.}~\bibnamefont {Kobus}}, \bibinfo {author} {\bibfnamefont {S.}~\bibnamefont {Kurdzia{\l{}}ek}},\ and\ \bibinfo {author} {\bibfnamefont {R.}~\bibnamefont {Demkowicz-Dobrza{\'{n}}ski}},\ }\href {https://doi.org/10.22331/q-2026-10-28-2251} {\bibfield  {journal} {\bibinfo  {journal} {{Quantum}}\ }\textbf {\bibinfo {volume} {10}},\ \bibinfo {pages} {2251} (\bibinfo {year} {2026}{\natexlab{b}})}\BibitemShut {NoStop}%
\end{thebibliography}%
\end{document}